\documentclass[a4paper,11pt]{article}
\usepackage{jheppub}
\usepackage{multirow}
\usepackage{physics}
\usepackage{booktabs}
\usepackage{tabularx}
\usepackage{graphicx} 
\usepackage{hyperref} 
\usepackage{amsmath}  
\usepackage{comment}
\usepackage{needspace} 
\usepackage{array}
\usepackage{footnote}

\title{\boldmath Probing the Higgs Self-Coupling with an XFEL Compton $\boldsymbol{\gamma\gamma}$ Collider at $\boldsymbol{\sqrt{s} = 380}$ GeV}

\author[a, b, 1]{Santiago Ampudia Castelazo,}
\author[a, b, 1]{Umar Sohail Qureshi,}
\note{Equal contribution.}
\author[b]{Tim Barklow,}
\author[b]{and Ariel Schwartzman}
\affiliation[a]{Department of Physics, Stanford University, Stanford, CA 94305, USA}
\affiliation[b]{SLAC National Accelerator Laboratory, Menlo Park, CA 94025, USA}

\emailAdd{sch@slac.stanford.edu}
\emailAdd{uqureshi@cern.ch}

\abstract{We present a study probing the Higgs self-coupling with the X-ray free-electron laser Compton $\gamma\gamma$ Collider (XCC) concept. The analysis is performed considering the $\gamma\gamma \rightarrow HH \rightarrow bb\overline{bb}$ channel and results are then extrapolated to obtain a projection on the Higgs self-coupling sensitivity that ranges between 7\% and 12\%. An ensemble of boosted decision trees is trained to discriminate between signal and backgrounds, paired with a genetic algorithm optimizer to combine the final classifier outputs. This study suggests that an X-ray FEL-based $\gamma\gamma$ collider is a powerful tool to probe the mechanism of electroweak symmetry breaking complementary to \(e^+e^-\) Higgs factories and future high energy hadron colliders.}

\begin{document}
\maketitle
\flushbottom

\section{Introduction}

The discovery of the Higgs boson at the Large Hadron Collider (LHC) in 2012 \cite{AadHiggs} marked a milestone in particle physics, confirming the mechanism of electroweak symmetry breaking, as described by the Standard Model (SM). Although the Higgs mass and its couplings to gauge bosons and fermions have been probed with increasing precision, one of the most critical parameters of the Higgs sector remains unmeasured: the Higgs self-coupling $\lambda_{HHH}$. This parameter governs the shape of the Higgs potential and is directly related to the mechanism of spontaneous symmetry breaking and the nature of the electroweak phase transition. A precise measurement of $\lambda_{HHH}$ is essential to fully reconstruct the Higgs potential and to test the SM predictions against scenarios of new physics, such as extended Higgs sectors, composite Higgs models, and electroweak baryogenesis~\cite{HiggsSymmetries,KanemuraElectroweak}.

The most direct way to access the Higgs self-coupling is through double-Higgs production at colliders. However, this process is rare and experimentally challenging due to small production cross sections and complex multi-jet final states. At the HL-LHC, the projected precision sensitivity to $\lambda_{HHH}$, assuming the SM value for the Higgs self-coupling, is $\sim30\%$ \cite{projectionsHLLHC}, far from the level 20\% that could allow one to detect deviations from the SM and directly probe the structure of extended Higgs sectors \cite{GuptaHiggs}. Hence, the Higgs self-coupling measurement is a major focus of future \(e^+e^-\) Higgs factories. The most recent update of the Higgs self-coupling projected precision with the ILD detector at a $550$~GeV linear collider facility is $11$\% for the SM value~\cite{ILD_HH}.

Photon colliders, based on Compton back-scattering of laser light off high-energy electron beams, provide a particularly attractive avenue for precision Higgs physics. 
Higgs pair production in  a $\gamma\gamma$ collider involves loop-level processes with direct sensitivity to $\lambda_{HHH}$, and benefits from enhanced signal-to-background separation compared to hadron collisions~\cite{Kawada:2012uy,Berger:2025ijd}. Previous concepts of $\gamma\gamma$ colliders, such as TESLA~\cite{TESLA}, CLICHE~\cite{CLICHE}, and SAPHIRE~\cite{SAPHIRE}, relied on optical laser systems, resulting in broad and asymmetric $\gamma\gamma$ energy spectra. These features limited the achievable precision and experimental feasibility of such proposals.

Recent advances in X-ray Free-Electron Laser (XFEL) technology have opened new possibilities for photon colliders. In particular, the X-ray Compton Collider (XCC) \cite{BarklowXCC} is a novel $\gamma\gamma$ collider concept that utilizes XFEL beams to produce Compton backscattered photons from electron beams of 62.6~GeV and 140-190~GeV, enabling $\gamma\gamma$ collisions at $\sqrt{s} = 125$~GeV for single-Higgs production and $\sqrt{s} = 280$~GeV or $\sqrt{s} = 380$~GeV for double-Higgs production. The use of soft X-ray photons leads to a sharply peaked and tunable $\gamma\gamma$ energy spectrum that enhances the precision of the Higgs measurements.

This study is the first physics simulation of the XCC photon-photon collider focused on measuring the Higgs self-coupling via the process $\gamma\gamma \rightarrow HH \rightarrow bb\overline{bb}$. Using a Delphes-based \cite{Delphes} detector simulation with an adapted version of the 2024 Silicon Detector (SiD) configuration \cite{DimitrisSiD}, a detailed analysis that accounts for all backgrounds is performed, including those originating from $\gamma\gamma$ processes, as well as those from residual $e\gamma$ and $e^+e^-$ interactions intrinsic to $\gamma\gamma$ colliders. Signal-background separation is optimized using multivariate techniques, and the statistical significance and expected precision on the double Higgs cross-section and the self-coupling parameter are evaluated.





\section{The XFEL Compton Photon-Photon Collider}

Previous \(\gamma\gamma\) collider designs utilized Compton back-scattering of optical-wavelength lasers off multi-GeV electron beams. These optical Compton collider (OCC) concepts offered access to processes uniquely available in \(\gamma\gamma\) collisions, including the loop-induced production of Higgs boson pairs. However, OCCs suffer from several intrinsic limitations: reduced luminosity at the desired energy peak and complicated precision measurements due to the broad and asymmetric \(\gamma\gamma\) energy spectrum, which arises from the Compton back-scattering process; altered scattering cross-sections and complexities in interaction dynamics caused by nonlinear QED effects, such as multiphoton absorption induced by high-intensity laser interactions; substantial beam-induced backgrounds, including electron-positron pairs and secondary photons, which interfere with event detection and degrade signal-to-background ratios. Consequently, detector performance at traditional OCCs was significantly constrained. For more information on OCC physics and limitations, see Refs.~\cite{TESLA,TelnovTESLA,TelnovProblems, BarklowXCC, TelnovProblems2}.

The X-ray FEL-based Compton Collider (XCC) represents a fundamental redesign of the \(\gamma\gamma\) collider concept, utilizing high-energy electron beams and X-ray free-electron laser pulses to generate high-luminosity, almost monochromatic photon collisions. The XCC consists of three primary components: a high-gradient electron linac, an X-ray free-electron laser (XFEL) line, and a set of interaction points where Compton scattering and \(\gamma\gamma\) collisions take place.

\begin{figure}[htbp]
    \centering
    \includegraphics[width=0.95\textwidth]{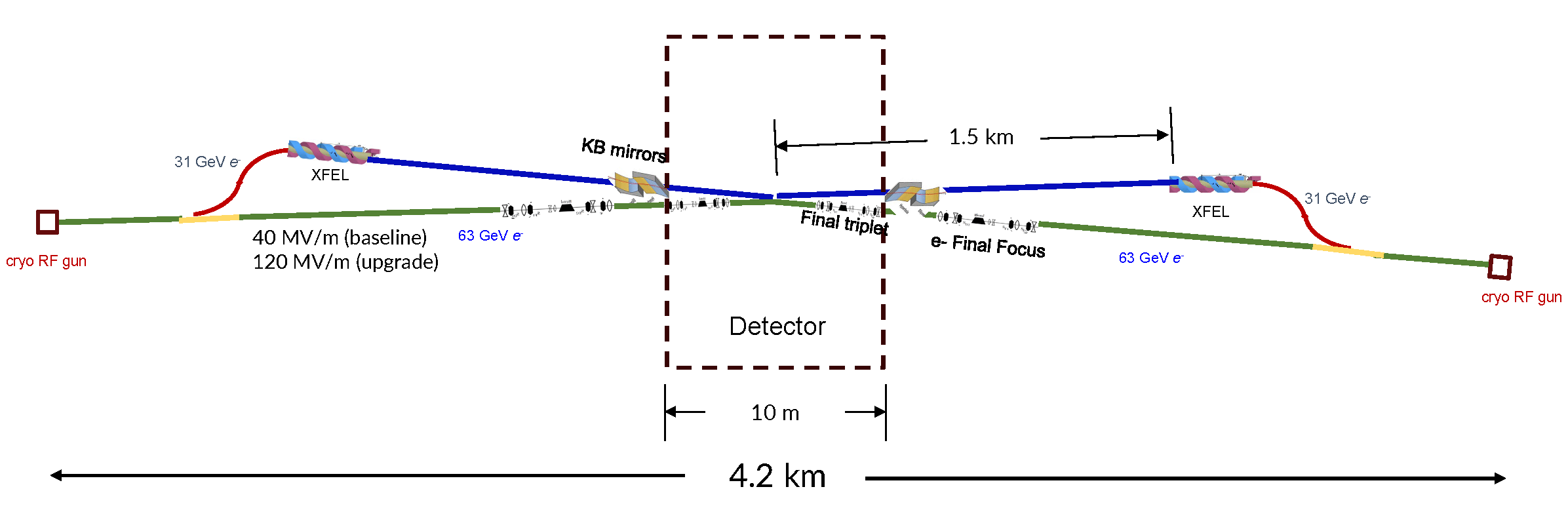}
    \caption{Schematic of XCC including cryogenic RF injector,  Linac, electron beam final focus (FF), and XFEL.}
    \label{fig:XCCSchematic}
\end{figure}

Polarized electron beams are generated by a cryogenic RF photo-injector and accelerated to 62.8~GeV (single Higgs) or 140-190~GeV (double Higgs) using C-band cold copper distributed coupling (C\textsuperscript{3}) linac structures. The beam is split mid-linac: every other electron bunch (at an energy of 30~GeV) is diverted to the XFEL line, where it passes through a helical undulator to produce circularly polarized 1~keV X-ray pulses. These pulses are subsequently focused to nanometer-scale waist sizes using Kirkpatrick-Baez mirrors. The remaining electron bunches continue down the linac to reach the full 62.8~GeV or 140-190~GeV energy. A schematic of XCC is shown in Figure~\ref{fig:XCCSchematic}.

At the Compton interaction point (IPC), these high-energy electron bunches collide with the counter-propagating XFEL pulses. Inverse Compton scattering occurs, generating photons with energies up to the full electron energy of 62.5~GeV or 140-190~GeV. The polarization and energy of the scattered photons are controlled through the helicity configurations of both the electrons and the laser photons. To maximize mono-chromaticity and suppress nonlinear QED effects, the design ensures a small laser intensity parameter \(\xi^2 \sim 0.1\).

\begin{figure}[htbp]
    \centering    \includegraphics[width=\linewidth]{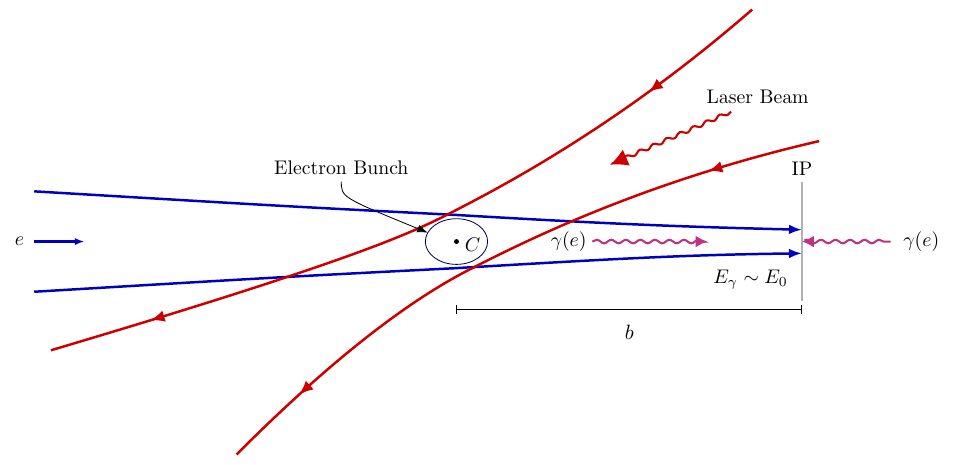}
    \caption{Close-up of the IP, illustrating the Compton process using only one side of the collider. The diagram can be mirrored to obtain the other side. }
    \label{fig:closeUpIP}
\end{figure}

The scattered photons continue downstream to the \(\gamma\gamma\) interaction point (IP), where two such photon beams are brought into collision. The design places the IPC and IP at a small distance (60~\(\mu\)m) apart to minimize transverse spreading due to the angular divergence of the scattered photons. Because only about 25\% of electrons convert to photons at the IPC, residual \(e^-\gamma\), and \(e^-e^-\) interactions also occur at the IP, and are taken into account in the overall luminosity and background simulations. Figure \ref{fig:closeUpIP} shows a close-up of the IP.

XCC's geometry supports a low crossing angle (2~mrad), with crab cavities implemented to rotate the electron beams for head-on collisions. A high solenoidal field (5~T) provides detector shielding and constrains low-angle background particles. Beam extraction downstream of the IP is enabled by a large-aperture final focus quadrupole (QD0), with energy deposition in QD0 minimized through strategic masking and beamline layout.

This multistage design allows for a clean and highly peaked \(\gamma\gamma\) energy spectrum, making XCC particularly suitable for precision studies at both the 125~GeV and 280-380~GeV CoM energy configurations, and could eventually be scaled to 10 TeV using wakefield electron acceleration. The ability to independently tune the electron beam and laser system, and the well-understood interactions at both the IPC and IP, provide a high degree of control over the collision environment, forming the foundation for the di-Higgs measurement program at \(\sqrt{s} = 280\)~GeV or \(\sqrt{s} = 380\)~GeV.

\section{\texorpdfstring{Luminosity Spectrum: XCC as a $\boldsymbol{\gamma\gamma}$, $\boldsymbol {e^+e^-} $ and $\boldsymbol {e\gamma}$  Collider}{}}

A key parameter describing \(\gamma\gamma\) collisions is $x = 4E_{e}w_{0}/m^2_e$, where $m_e$ is the electron mass, $w_{0}$ is the laser wavelength, and $E_e$ is the electron beam energy. The maximum Compton photon energy is given by $\frac{x}{(x+1)}E_e$. For increasing values of $x$, the high-energy photon spectrum becomes more peaked towards the maximum energy that becomes closer to $E_e$. Since the photons produced in the peak region are the most important for physics, enhancing the physics potential of \(\gamma\gamma\) colliders calls for operation at higher values of $x$. However, larger $x$
values are problematic due to the linear QED thresholds of $x=4.82$ and $x=8$ for the processes $\gamma \gamma_{0} \rightarrow e^+e^-$ and $e^- \gamma_{0} \rightarrow e^-e^+e^-$ respectively, where $\gamma$ and $\gamma_0$ refer to the Compton-scattered and laser photon respectively. For this reason, all previous 
\(\gamma\gamma\) collider concepts limited their operation to  $x < 4.8$ and infrared/optical laser frequencies to limit the creation of electron-positron pairs. 

However, the emergence of X-ray FELs has enabled the possibility of operating \(\gamma\gamma\) colliders at extremely high values of $x$ ($x \ge 1000$) such that the impact of the linear QED
process creating $e^+e^-$ pairs is overcome by a \(\gamma\gamma\) luminosity distribution with respect to the center-of-mass (CoM) energy that is extremely sharply peaked near the maximum CoM value. Furthermore, by setting the electron helicity  $\lambda_e$ and the photon circular polarization $P_{c}$ such that $\lambda_e P_{c}>0$, the rate for $\gamma \gamma_{0} \rightarrow e^+e^-$ is reduced through helicity suppression, leading to an overall higher \(\gamma\gamma\) luminosity. 

These characteristics bring the \(\gamma\gamma\) luminosity distribution more in line with that of $e^+e^-$ colliders, increasing the signal-to-background ratio of narrow resonances and enabling kinematic fitting of final-state momenta with full energy-momentum constraints. X-ray FELs have another advantage in that they
circumvent the timing and repetition rate problems associated with high-power optical lasers. 

\begin{figure}[htbp]
    \centering    \includegraphics[width=0.75\textwidth]{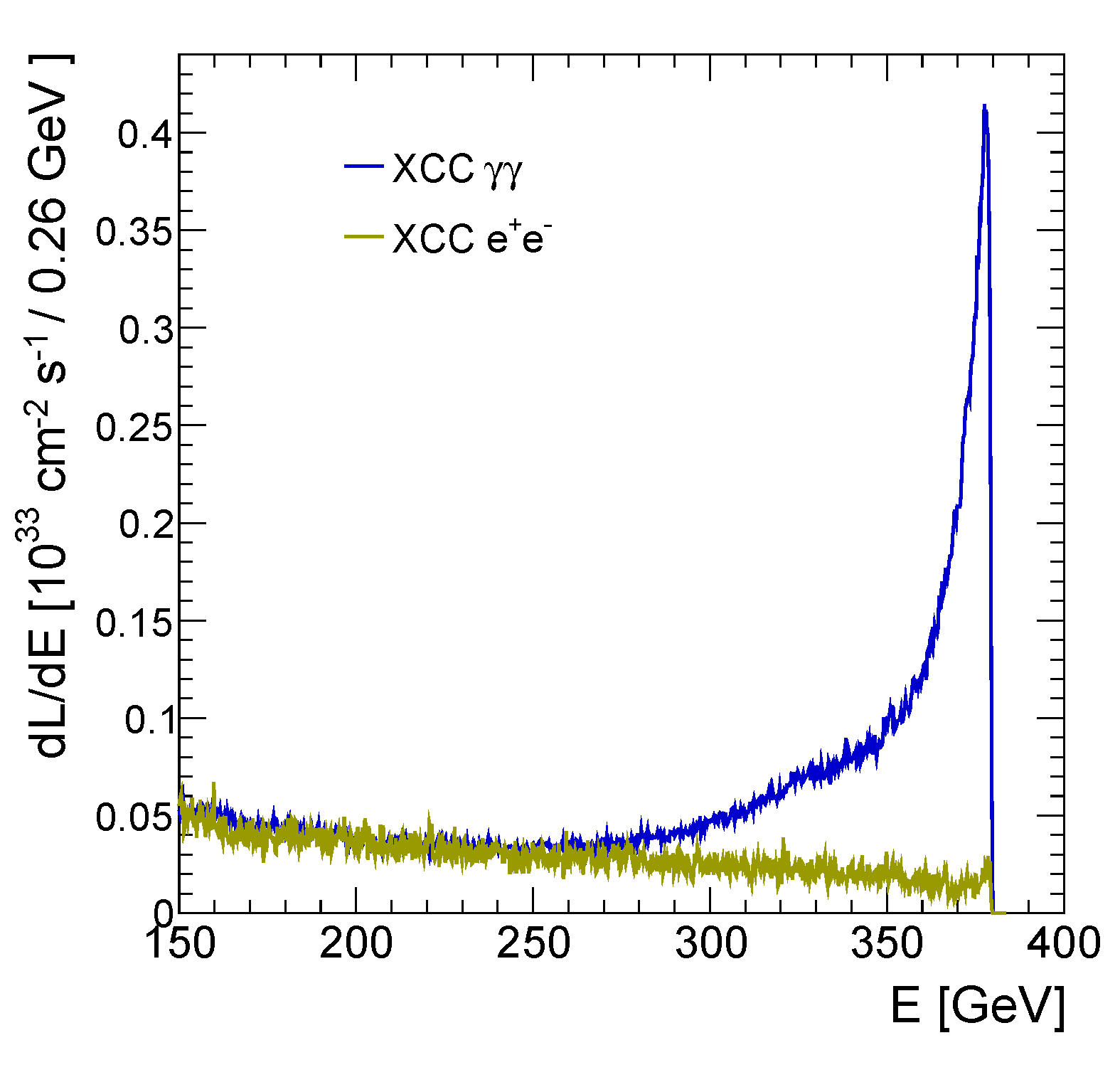}
    \caption{Luminosity spectra for the two most important initial state particle combinations: $\gamma\gamma$ collisions (blue) and  $e^+e^-$ collisions (gold) for the center-of-mass energy range $150\ \mathrm{GeV} < \sqrt{\widehat{s}} < 380\ \mathrm{ GeV}$. Events due to $e^-\gamma$ and $e^-e^-$ collisions make up only 1.5\% of the total background after the final signal-background separation.}  
    \label{fig:lumiHH}
\end{figure}

The XCC \(\gamma\gamma\) collider concept is designed to operate at a value of $x=1000$, leading to the sharp $\gamma\gamma$ luminosity spectrum shown in Figure~\ref{fig:lumiHH}.

The luminosity spectrum has been obtained using the \textsc{Cain} Monte Carlo code~\cite{CAIN}, which includes a 0.1\% electron beam energy spread, linear QED Bethe–Heitler scattering, and non-linear QED effects in Compton and Breit–Wheeler scattering.

\begin{figure}[htbp]
    \centering
    \includegraphics[width=0.8\textwidth]{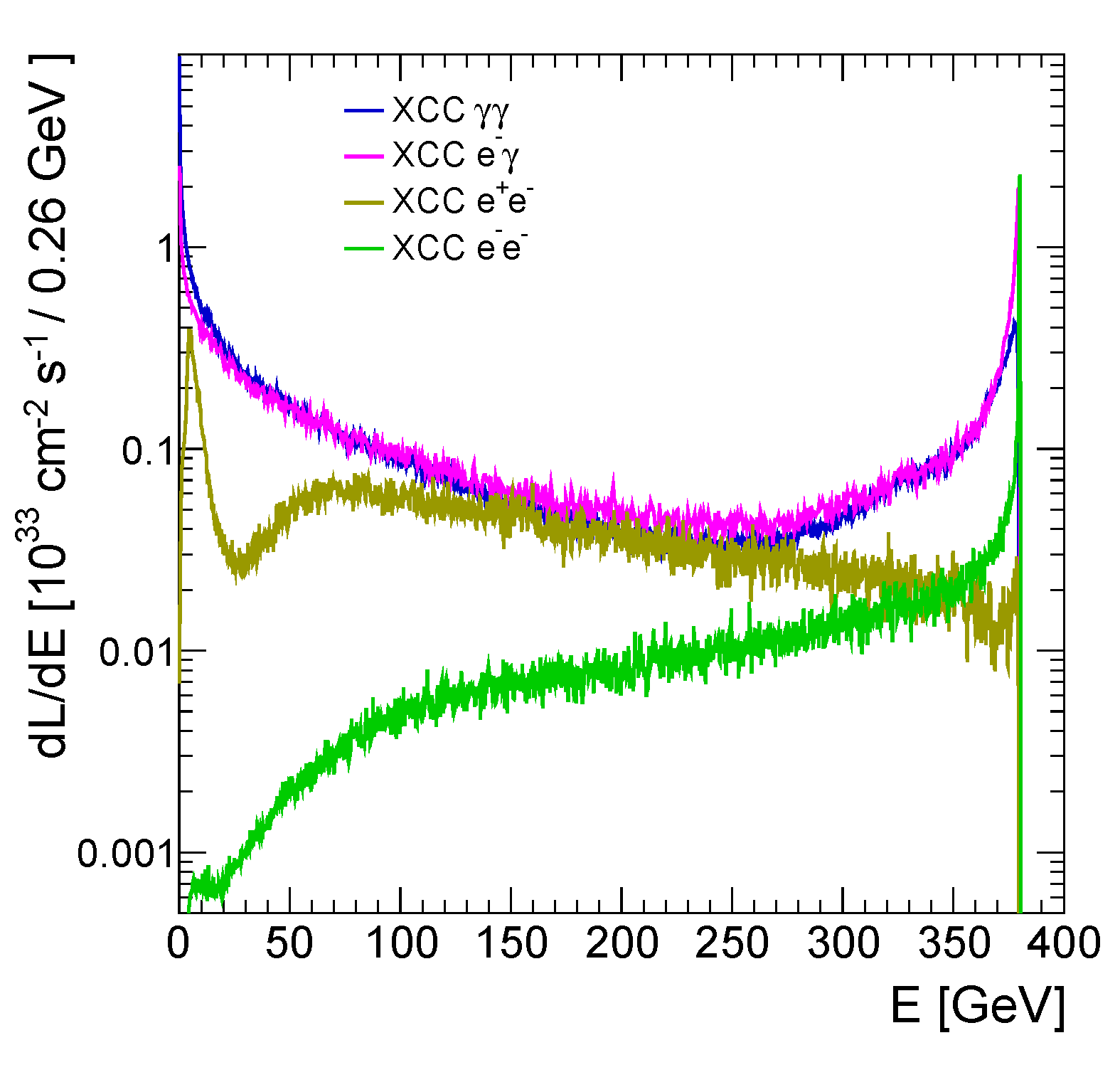}
    \caption{Luminosity spectrum for collisions of  $\gamma\gamma$  (blue), $e^-\gamma$ (red) $e^+e^-$ (gold) and $e^-e^-$ (green) for the full center-of-mass energy range $0\ \mathrm{GeV} < \sqrt{\widehat{s}} < 380\ \mathrm{ GeV}$.}
    \label{fig:eGammaSpectrumFull}
\end{figure}

In addition to \(\gamma\gamma\) 
collisions, and given that the Compton conversion efficiency at the IPC is roughly 25\%, unconverted electrons from one beam will collide with high-energy Compton-scattered photons, electrons and positrons from the opposing beam creating \( e^-\gamma\), \( e^-e^- \) 
and \(e^+e^-\) collisions at the IP. 
Figure~\ref{fig:eGammaSpectrumFull} shows the luminosity spectra for \(\gamma\gamma\), \( e\gamma \), and $ee$ collisions at the XCC.

\section{Higgs Self-coupling Measurement at an XFEL Photon-Photon Collider}

A \(\gamma\gamma\) collider presents multiple advantages and points of complementarity to $e^+e^-$ for probing the Higgs self-coupling. First, at $\sqrt{s} \approx 380$~GeV, the loop-induced di-Higgs process via intermediate \( W \) bosons and top quarks reaches its maximum cross-section, which is approximately twice that of \( e^+e^- \to ZHH \) at 500 GeV~\cite{BarklowXCC}. The enhancement arises despite \( \gamma\gamma \to HH \) being a one-loop process, due to the absence of phase-space suppression from an associated final-state boson. This enables the measurement of the trilinear coupling at a lower CoM energy compared to an $e^+e^-$ collider.  The lower beam energy requirement directly translates into a more compact and cost-efficient collider infrastructure. Operating at 280-380~GeV, it eliminates the need for a longer linac and reduces the overall energy budget. This lower energy configuration, in addition to the absence of positrons, can reduce power consumption and construction complexity, potentially leading to significant cost reductions. See Section 1.5 of Ref.~\cite{BarklowXCC} for a detailed breakdown of the expected cost of XCC.

\begin{figure}[htbp]
    \centering
    \begin{minipage}{0.465\linewidth}
    \centering
        \includegraphics[width=\linewidth]{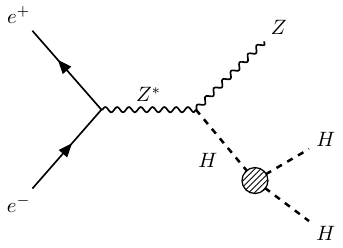}
    \end{minipage} 
 \begin{minipage}{0.495\linewidth}
    \centering
        \includegraphics[width=\linewidth]{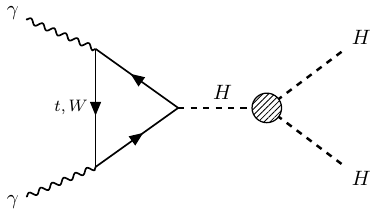}
    \end{minipage}  

    \caption{Feynman diagrams for di-Higgs production at XCC at \( \sqrt{s} = 380 \) GeV and linear \( e^+e^- \) colliders at \( \sqrt{s} = 500 \) GeV. Left: Tree-level \( e^+e^- \to ZHH \) process typical of linear colliders such as the ILC, which includes an on-shell \( Z \) boson in the final state. Right: Loop-level \( \gamma\gamma \to HH \) process at the XCC, mediated by virtual \( W \) bosons and top quarks. }
    \label{fig:productionMechanism}
\end{figure}

\begin{figure}[htbp]
    \centering
    \includegraphics[width=0.8\linewidth]{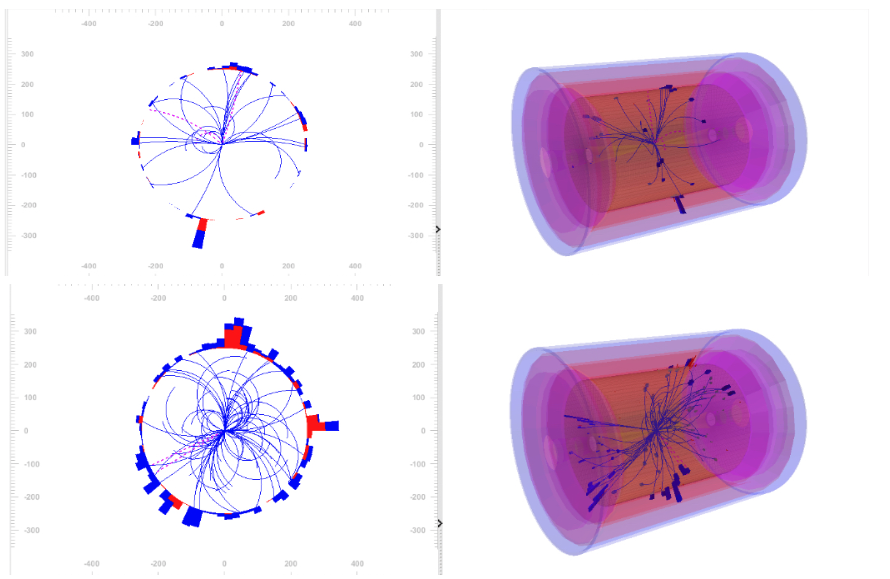}
    \caption{Event displays. Top: $\gamma\gamma \rightarrow HH \rightarrow bb\overline{bb}$ at $\sqrt{s}$=380\,GeV. Bottom: $e^+e^- \rightarrow ZHH \rightarrow qqbb\overline{bb}$ at $\sqrt{s}$=550\,GeV.}
    \label{fig:eventDisplaysProduction}
\end{figure}

Second, Higgs pair production at a \(\gamma\gamma\) collider gives rise to a simpler final state consisting of only four jets in the $H \rightarrow b \overline{b} $ channel. This is because the process \( \gamma\gamma \to HH \) proceeds primarily through one-loop diagrams involving virtual \( W \) bosons and top quarks, which yield a final state composed solely of the Higgs boson pair with no accompanying vector boson. In contrast, \( e^+e^- \to ZHH \) proceeds via a tree-level diagram that necessarily includes an on-shell \( Z \) boson in the final state, as shown in Figure~\ref{fig:productionMechanism}. As illustrated by representative event displays in Figure~\ref{fig:eventDisplaysProduction}, this fundamental difference in production topology leads to a substantially cleaner experimental signature at XCC, with fewer final-state particles, lower hadronic activity, and reduced combinatorial ambiguity in event reconstruction. This generally translates into a more accurate event reconstruction; therefore, it allows for better signal-background discrimination and enhanced measurement sensitivity.

\begin{figure}[htbp]
    \centering
    \includegraphics[width=0.7\textwidth]{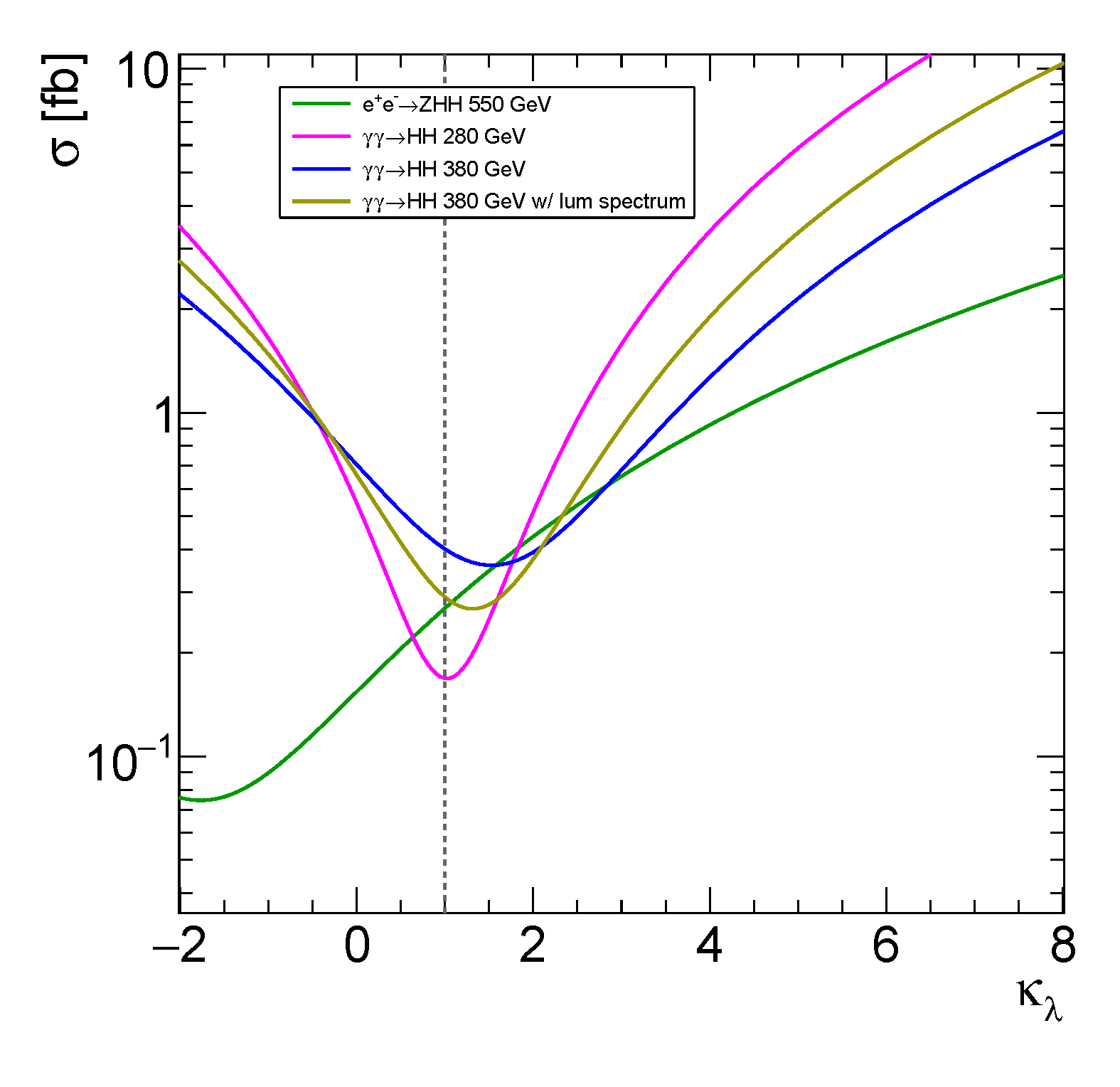}
    \caption{Cross-section versus $\kappa_\lambda$ for $\gamma\gamma\rightarrow HH$ at $\sqrt{s} = 380$~GeV (blue curve), $\gamma\gamma\rightarrow HH$ at $\sqrt{s} = 280$~GeV (magenta curve) and $e^+e^-\rightarrow ZHH $ at $\sqrt{s} = 550$~GeV (green curve) Also shown is the convolution of the $\gamma\gamma\rightarrow HH$ cross-section with 
    a normalized $\gamma\gamma$ luminosity spectrum  for $250 < \widehat{s} < 380$~GeV  (gold curve)~\cite{Berger:2025ijd}.}
    \label{fig:sigHH_vs_kLam}
\end{figure}

The dependence of the $\gamma\gamma\rightarrow HH$ cross-section on $\kappa_\lambda$ for $\gamma\gamma\rightarrow HH$ at $\sqrt{s} = 280\ \textrm{and}\ 380$~GeV, and for $e^+e^-\rightarrow ZHH $ at $\sqrt{s} = 550$~GeV is shown in Figure~\ref{fig:sigHH_vs_kLam}. 
It is interesting to note that a \( \gamma\gamma\) collider has the strongest  cross-section dependence on $\kappa_\lambda$ for $\kappa_\lambda \ne 1$.
This means that a \( \gamma\gamma \to HH \) collider is particularly sensitive to BSM models that lead to deviations of the Higgs self-coupling from its SM value. Furthermore, this dependence is different from that of proton-proton and $e^+e^-$ colliders, providing a complementary probe of the Higgs potential. 

\section{Monte Carlo Samples and Event Reconstruction}

\subsection{Event Generation}
\label{sec:ev_gen}
The simulation chain of the XFEL \( \gamma\gamma\) collider starts with the \textsc{Cain} \cite{CAIN} Monte Carlo package. In particular, \textsc{Cain} is used to simulate the initial beam-beam and beam-laser interactions and non-linear QED effects for \(\gamma\gamma\), \( e^+e^- \) and \( e\gamma \) processes\footnote{Even though the XCC has significant $e^-e^-$ luminosity, no $e^-e^-$ processes contributed were found to contribute meaningfully to this analysis.}. Events are generated using an energy spread ranging from $\sqrt{s} = 380$~GeV and an integrated luminosity of $\mathcal{L} = 4900~\text{fb}^{-1}$ for a 10-year run-time, leveraging the sharply peaked luminosity spectrum characteristic of the XCC. The energy spread accounts for the energy loss from the Compton backscattering process at the IPC, where only a fraction of the electron energy is transferred to each photon depending on the scattering angle, resulting in a broad photon energy spectrum that is further modified by beamstrahlung and radiative losses before reaching the IP. The output of \textsc{Cain} containing the generated photon, electron, and positron luminosity spectra is then interfaced with \textsc{Whizard}~\cite{WHIZARD} for event generation. Generated events are then filtered\footnote{This requirement is a safe assumption since the number of events with no $b$-quarks that would pass the preselection detailed in Sec.~\ref{analysis} is negligible. In earlier iterations of the study, samples were generated  without the filter and the assumption was verified.} with the requirement $N_{b\text{-quark}} \ge 1$  or $N_{c\text{-quark}} \ge 3$. All processes have the additional filter $m_{\text{hadrons}} > 250~\mathrm{GeV}$ at the parton level.

\subsection{Signal and Background Processes}
Signal events consist of $\gamma\gamma \rightarrow HH \rightarrow bb\overline{bb}$. The total $\gamma\gamma \rightarrow HH$ production cross-section at $\sqrt{s} = 380$~GeV is approximately 0.4~fb, but given that the events were generated using an energy spread, the production cross section of the signal in this study is about 0.37, leading to approximately 1812 expected $\gamma\gamma \rightarrow HH$ events.

A total of twelve relevant physical background processes were included in the analysis. These backgrounds were generated using the same simulation methodology as the signal. Due to the nature of the XCC environment near the IP, background processes are categorized into three types of interaction: gamma-gamma ($\gamma\gamma$), electron-gamma ($e\gamma$) and electron-positron ($e^+e^-$). Table \ref{tab:backgrounds} contains the expected number of events for each background. 

\begin{table}[htbp]
    \centering
    \begin{tabular}{ll}
        \hline
        \textbf{Process} & \textbf{Expected Events} \\
        \hline
        $\gamma\gamma \rightarrow HH\rightarrow bb\overline{bb}$           & 1,812 \\
        $\gamma\gamma \rightarrow W^+W^-\rightarrow q\overline{q}q\overline{q}$       & 3,813,000 \\
        $\gamma\gamma \rightarrow t\overline{t}$                                 & 2,866,000 \\
        $\gamma\gamma \rightarrow ZZ\rightarrow q\overline{q}q\overline{q}$           & 1,378,000 \\
        $\gamma\gamma \rightarrow q\overline{q}$                                 & 307,700 \\
        $\gamma\gamma \rightarrow ZH\rightarrow q\overline{q}H$                  & 8,202 \\
        \hline
        $e\gamma \rightarrow eq\overline{q}\textrm{ or }\nu q\overline{q}$            & 41,195 \\
        $e\gamma \rightarrow eq\overline{q}q\overline{q}\textrm{ or }\nu q\overline{q}q\overline{q}$ & 7,681 \\
        $e\gamma \rightarrow eq\overline{q}H$                                    & 3,282 \\
        \hline
        $e^+e^- \rightarrow q\overline{q}$                                       & 753,615 \\
        $e^+e^- \rightarrow W^+W^- \textrm{ or } ZZ\rightarrow q\overline{q}q\overline{q}$ & 152,212 \\
        $e^+e^- \rightarrow ZH \rightarrow q\overline{q}H$                       & 123,698 \\
        $e^+e^- \rightarrow t\overline{t}$                                       & 57,001 \\
        \hline
        \textbf{Total}                                                     & \textbf{9,513,398} \\
        \hline
    \end{tabular}
    \caption{List of background processes included in the analysis and their expected number of events for an integrated luminosity of $\mathcal{L} = 4900~\text{fb}^{-1}$ corresponding to a 10-year run-time at $\sqrt{s} = 380$~GeV.}
    \label{tab:backgrounds}
\end{table}

\subsection{Detector Simulation and Event Reconstruction}

\textsc{Delphes} was used to perform a fast detector simulation on the \textsc{Whizard} output files. A configuration card for XCC was developed based on that of the SiD detector for ILC \cite{PotterSiD,detectorsILC}.  The XCC, however, suffers from increased incoherent $e^+e^-$ pair production (IPP) from Bethe–Heitler $\gamma\gamma^* \to e^+e^-$, Breit–Wheeler $\gamma \gamma \to e^+e^-$, and Landau–Lifshitz $\gamma^*\gamma^* \to e^+e^-$ processes compared to its $e^+e^-$ counterparts. To compensate, the radius of the inner layer of the vertex detector, which is a crucial specification for flavor tagging, is increased from the SiD value of 1.4~cm to 1.7~cm. For this change, we leverage the \textsc{DetectorGeometry} and \textsc{TrackCovariance} modules in \textsc{Delphes}; the former takes as input, a geometric description of the detector's tracking system and the latter, using the geometry, provides an estimate for the track parameters, and associated covariance matrix, for each charged particle. A detailed description of these modules and their technical implementation is provided in Ref.~\cite{Bedeschi:2022rnj}.

The $|\cos\theta|<0.95$ angular coverage of the XCC vertex detector is also markedly reduced with respect to SiD's $|\cos\theta|<0.98$. The $p_T$ vs. $\theta$ distribution of charged particles from IPP determines the inner radius of the vertex detector at XCC as it does  at $e^+e^-$ colliders. \textsc{Cain} simulations demonstrate that the 1.7~cm radius is well above the minimum required as $< 0.01$\% of IPP charged particles  strike a 1.4~cm radius cylindrical beampipe
with angular extent $\abs{\cos\theta}<0.95$, 
which can be compared to 0.1\% at a 250 GeV CoM energy $e^+e^-$ linear collider with a 1.4~cm radius, $|\cos\theta|<0.98$ beampipe 
\cite{Ntounis:2025stc}. We accordingly use degraded numbers from the flavor tagging configuration described in Ref.~\cite{DimitrisSiD} for this study. In particular, we select a working point of the ParticleNet \cite{Qu:2019gqs} algorithm that gives an 85\% efficiency for $b$-jets with a mistag rate of 5\% for $c$-jets and 1\% for light jets. It is worth noting that this is a very conservative assumption: preliminary flavor tagging studies reveal that even with our modified detector, we observe significantly improved performance than the working point used in this study --- which assumes aggressive degradation, comparable with current LHC figures \cite{ATLAS:2025dkv, CMS-DP-2022-050}, as a worst-case scenario. Flavor tagging will be investigated again in  future XCC studies with both fast and full detector simulation.

\begin{figure}[htbp]
    \centering
    \includegraphics[width=0.49\textwidth]{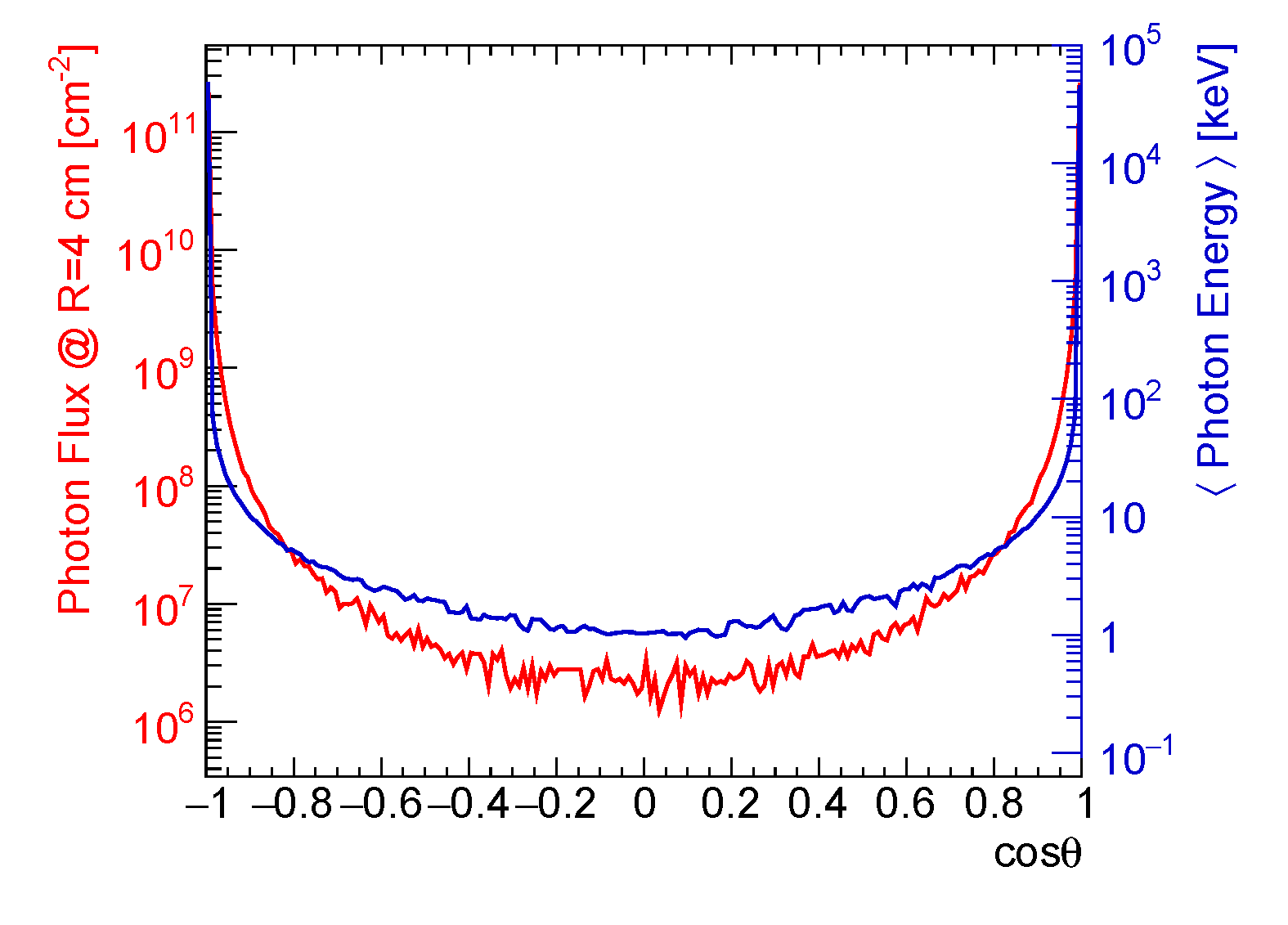}
    \includegraphics[width=0.49\textwidth]{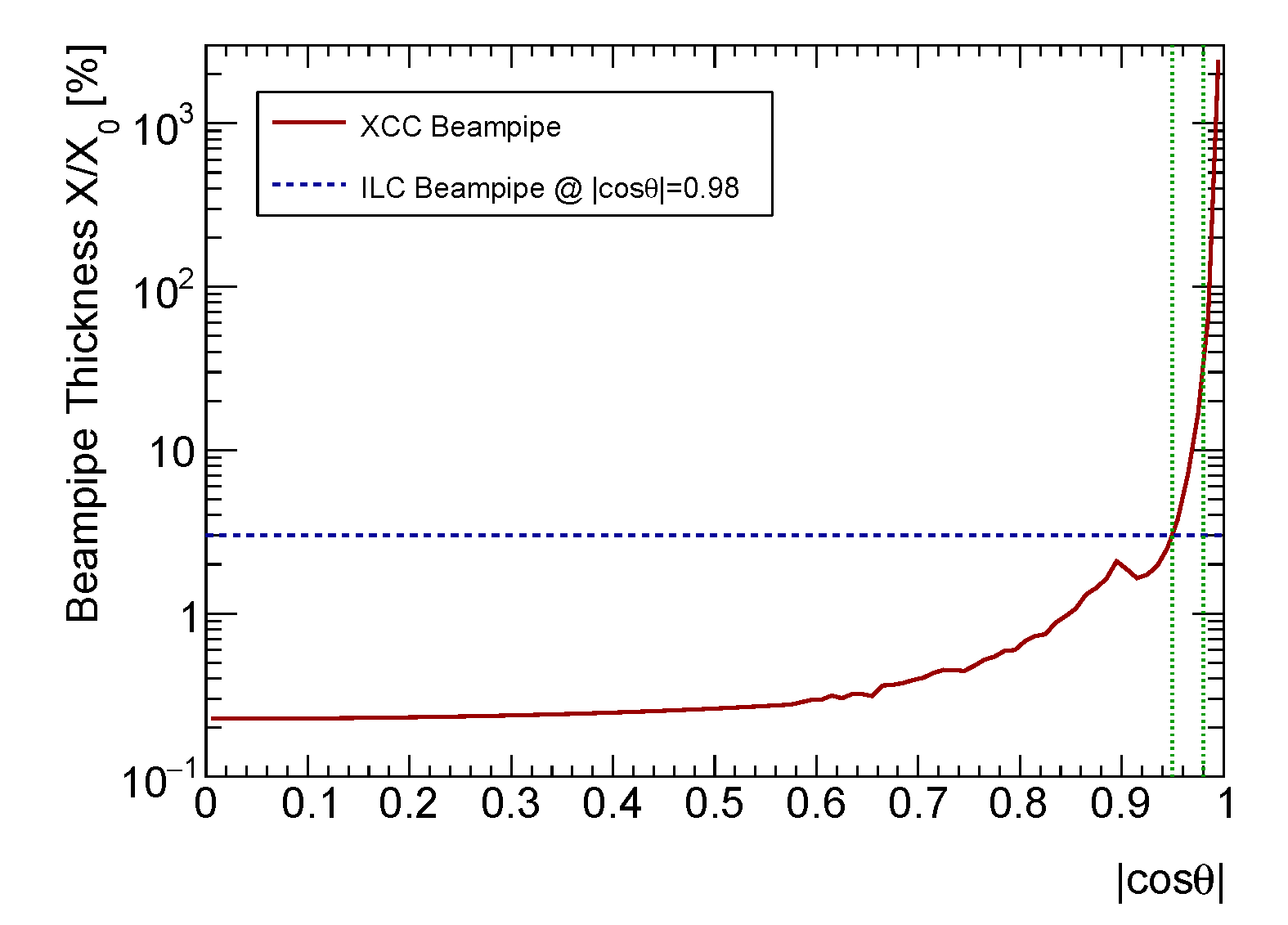}    
    \caption{Compton photons flux and mean photon energy versus $\cos\theta$ (left) and thickness of a gold-coated beryllium beampipe versus $\cos\theta$ (right), where $\theta$ is the angle of the photon with respect to the beam axis. The beampipe thickness is chosen to limit the X-ray background to  60 hits/cm$^2$  in a 100 $\mu\textrm{m}$ thick Si layer at $R=4$ cm.} 
    \label{fig:uniquePhotonBacks}
\end{figure}

A unique photon background is generated in the Compton collisions of XCC, as shown in the left-hand plot of Fig.~\ref{fig:uniquePhotonBacks}. There is a moderate flux of soft X-rays in the central region $|\cos\theta| < 0.85$ which can be mitigated with absorbers of $0.2\% - 1. 0\%$ $X_0$ (right-hand plot of Fig.~\ref{fig:uniquePhotonBacks}). The forward regions experience a steep increase in both the number and energy of background X-rays. This requires denser absorbers (up to $3.0\%$ $X_0$ at $|\cos\theta| = 0.95$) and poses significant challenges for detector design in the region $0.95 < |\cos\theta| < 0.99$, with essentially no instrumentation possible beyond $|\cos\theta| > 0.99$. Given these considerations, \textsc{Delphes}
particle flow objects (PFOs) in the very forward regions of the detector, specifically $|\cos\theta|>0.95$, are excluded from the analysis.  Again, this assumption is very conservative; in practice, dedicated detector elements and  background suppression algorithms are expected to be able to extend the reconstruction of physics objects to the $0.95 < |\cos\theta| < 0.98$ range.

Pileup in the form of multiple $\gamma\gamma\rightarrow \textrm{hadrons}$ events per bunch crossing is present at XCC, as it is at $e^+e^-$ linear colliders.  The total cross section
for $\gamma\gamma\rightarrow \textrm{hadrons}$ increases slowly from $\sqrt{\widehat{s}}=0.3$~GeV to 400~GeV with
a mean value of about 0.4~$\mu$b.  The average number of pileup events is calculated by convolving the $\gamma\gamma$, $e^-\gamma$ and $e^-e^-$ luminosity spectra with the total $\gamma\gamma\rightarrow \textrm{hadrons}$ cross-section; the spectra from $e^-$ beams has been included to simulate hadron production from virtual photon collisions with virtual and real photons.  The result is 9.1, 6.2, and 1.1 pileup events per bunch crossing for $\sqrt{s}=380, 280,$ and 125~GeV, respectively.  The number of pileup events is much less than at LHC and, for XCC $\sqrt{s}=280-380$~GeV, the same as a 3~TeV $e^+e^-$ linear collider.  For XCC at $\sqrt{s}=125$~GeV the number of pileup events is the same at ILC with $\sqrt{s}=500$~GeV.  Pileup events were not included in the simulated event samples used for this analysis as most of their PFOs are produced outside the $|\cos\theta|<0.95$ detector volume. Furthermore, preliminary pileup mitigation studies with dedicated machine learning algorithms (see Appendix \ref{sec:pileup}), reveal that this pileup can be effectively suppressed with negligible impact on the jet energy resolution and thus the analysis sensitivity. Future studies with full detector simulation covering $|\cos\theta|<0.99$ will include pileup.

\begin{figure}
    \centering   
    \begin{minipage}{0.495\linewidth}
    \centering
        \includegraphics[width=\linewidth]{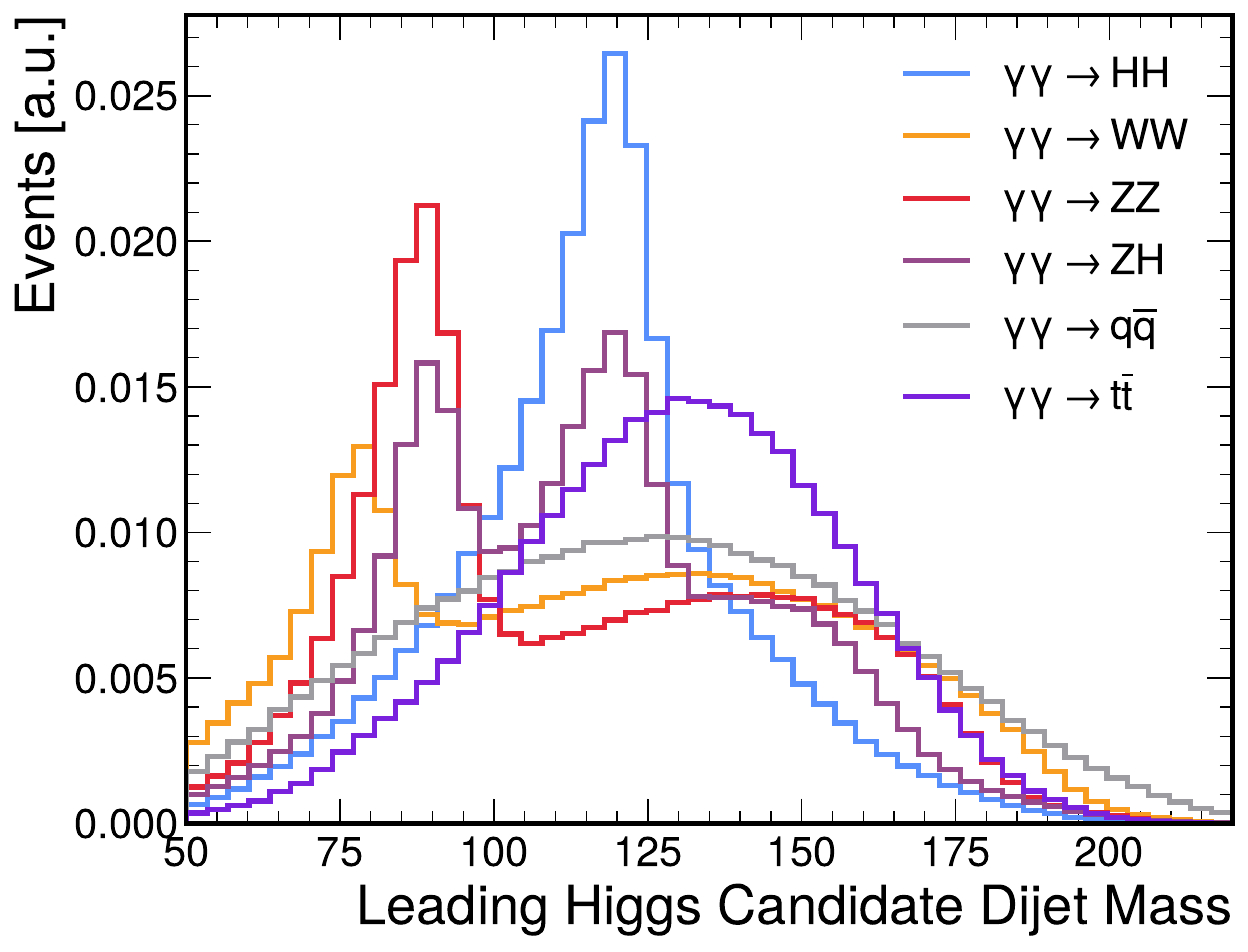}
    \end{minipage}
    \begin{minipage}{0.495\linewidth}
    \centering
        \includegraphics[width=\linewidth]{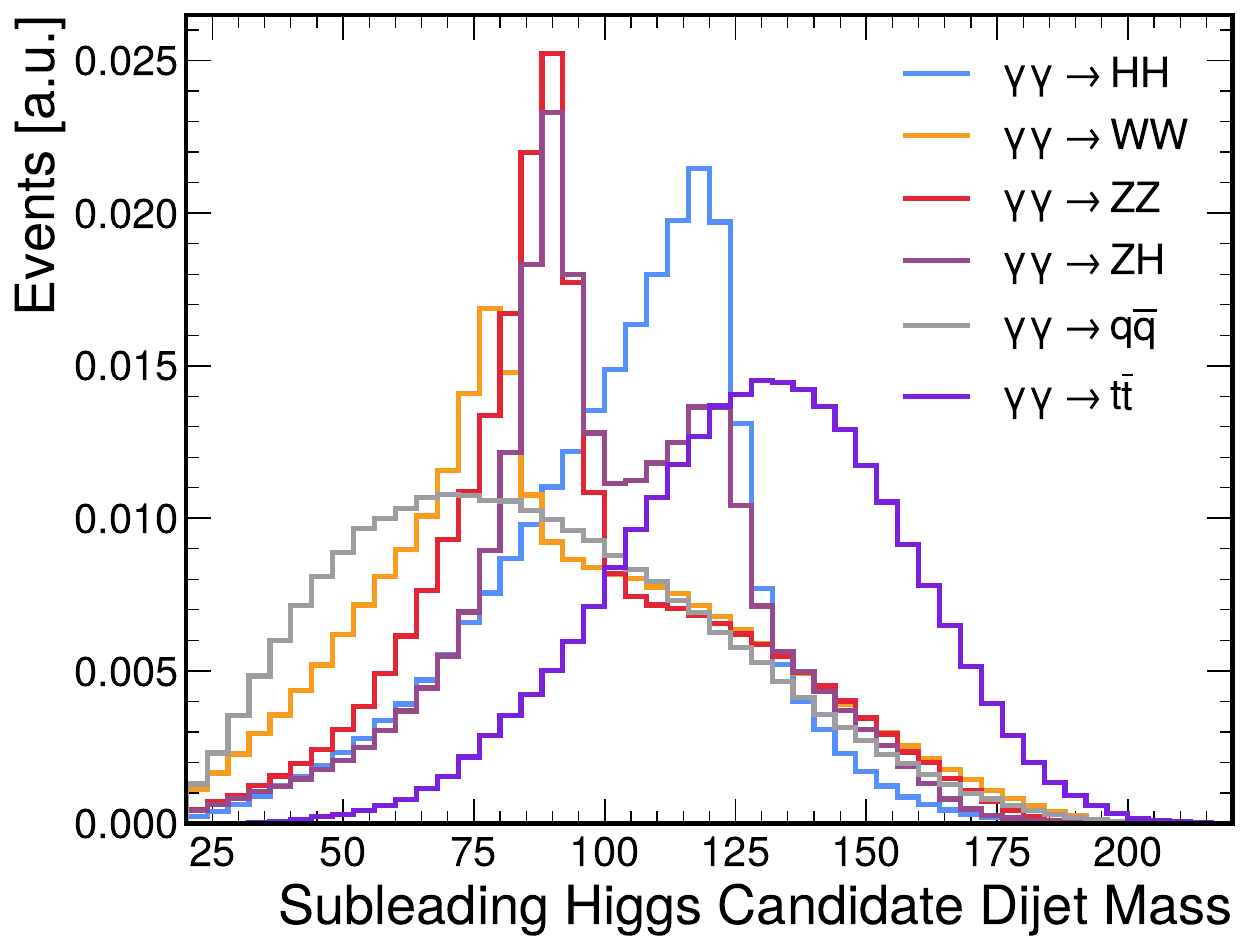}
    \end{minipage}
    \begin{minipage}{0.495\linewidth}
    \centering
        \includegraphics[width=\linewidth]{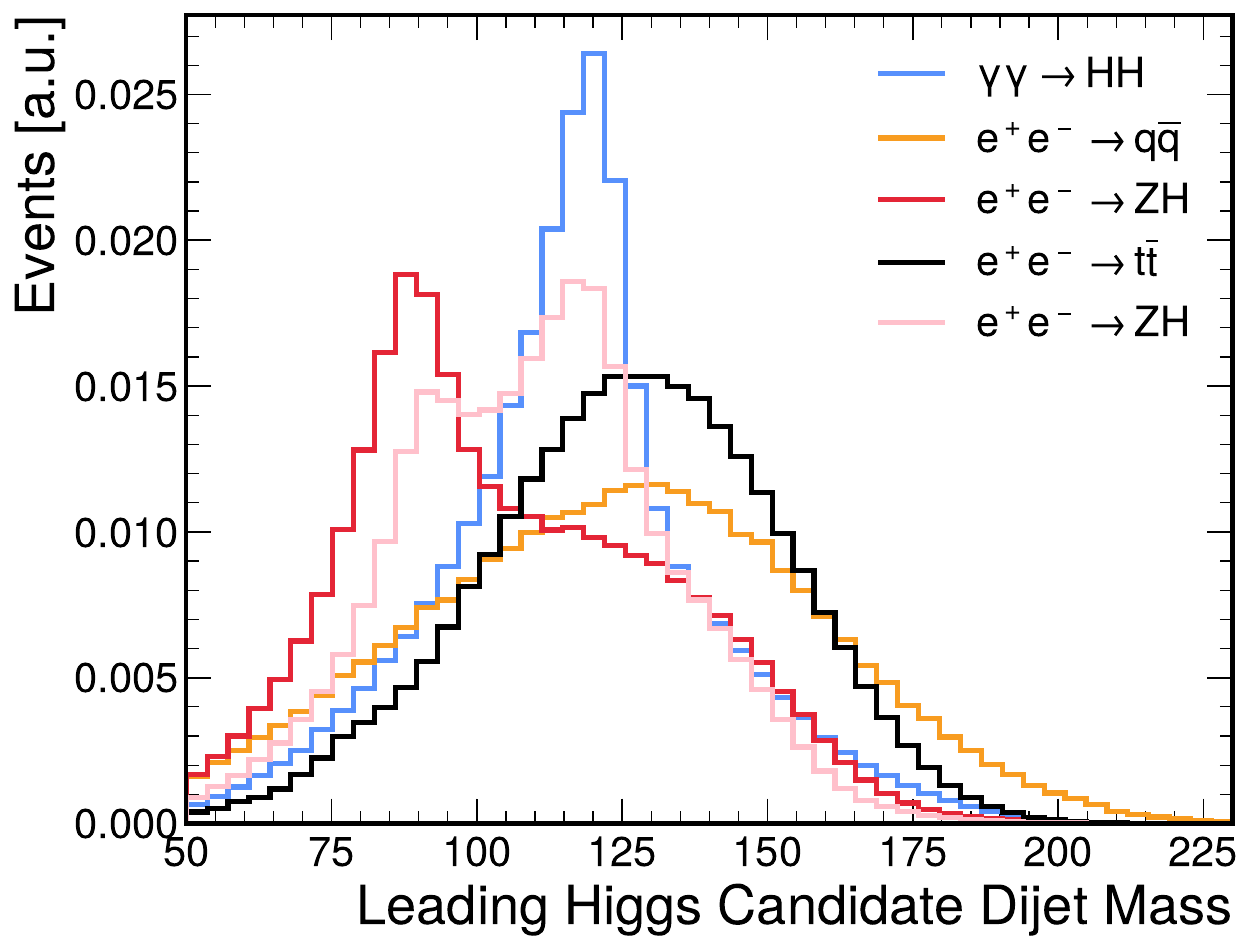}
    \end{minipage}
    \begin{minipage}{0.495\linewidth}
    \centering
        \includegraphics[width=\linewidth]{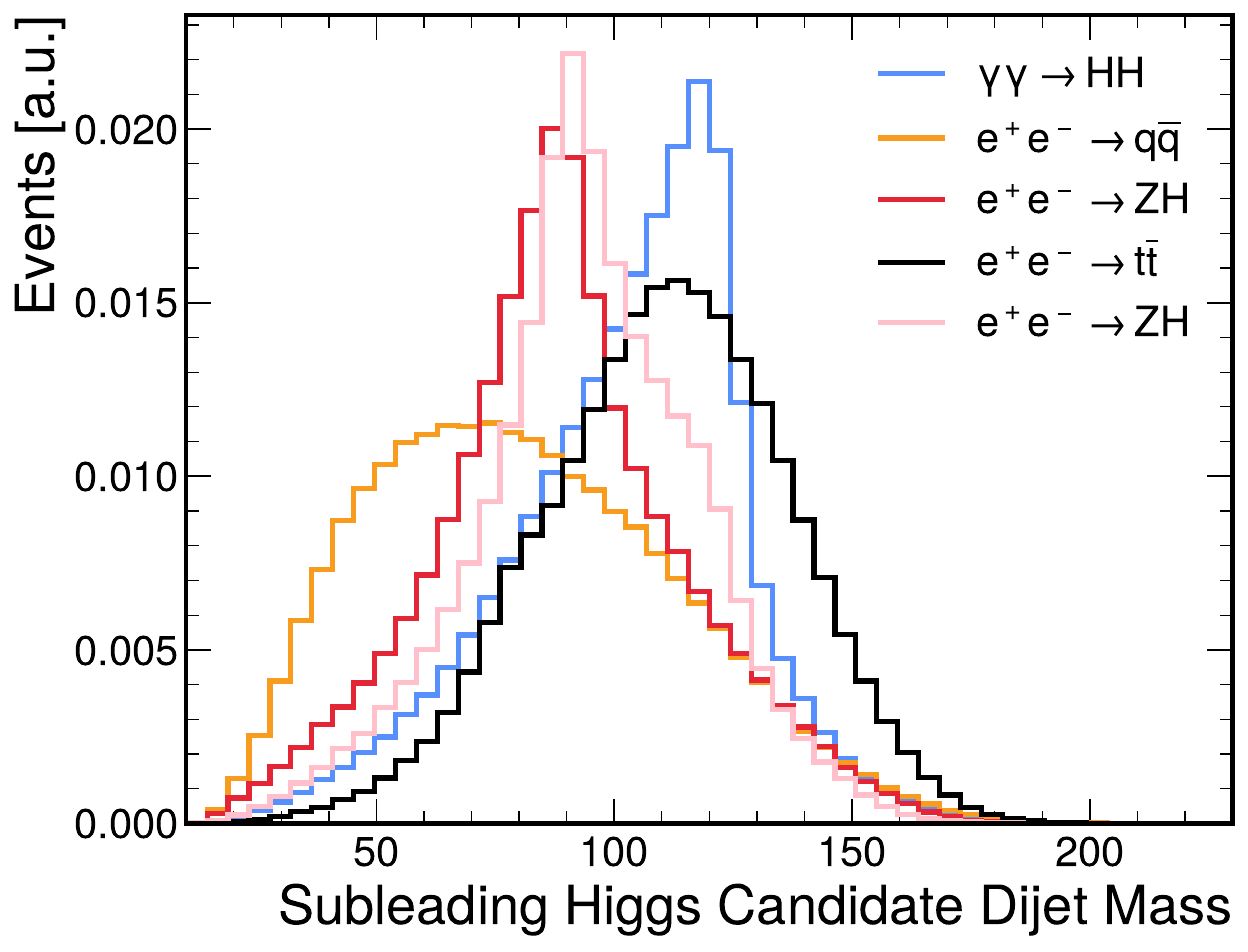}
    \end{minipage}
    \begin{minipage}{0.495\linewidth}
    \centering
        \includegraphics[width=\linewidth]{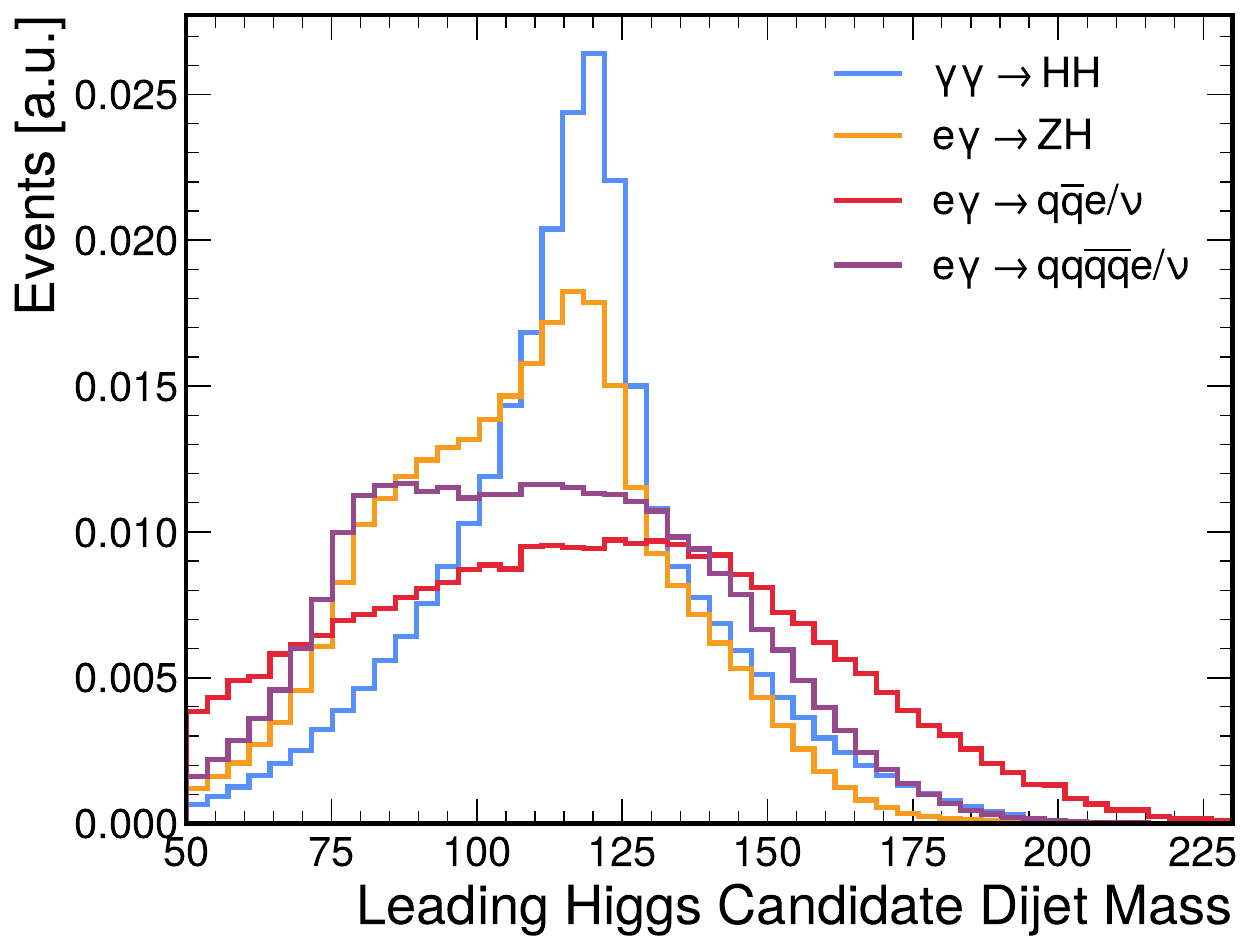}
    \end{minipage}
    \begin{minipage}{0.495\linewidth}
    \centering
        \includegraphics[width=\linewidth]{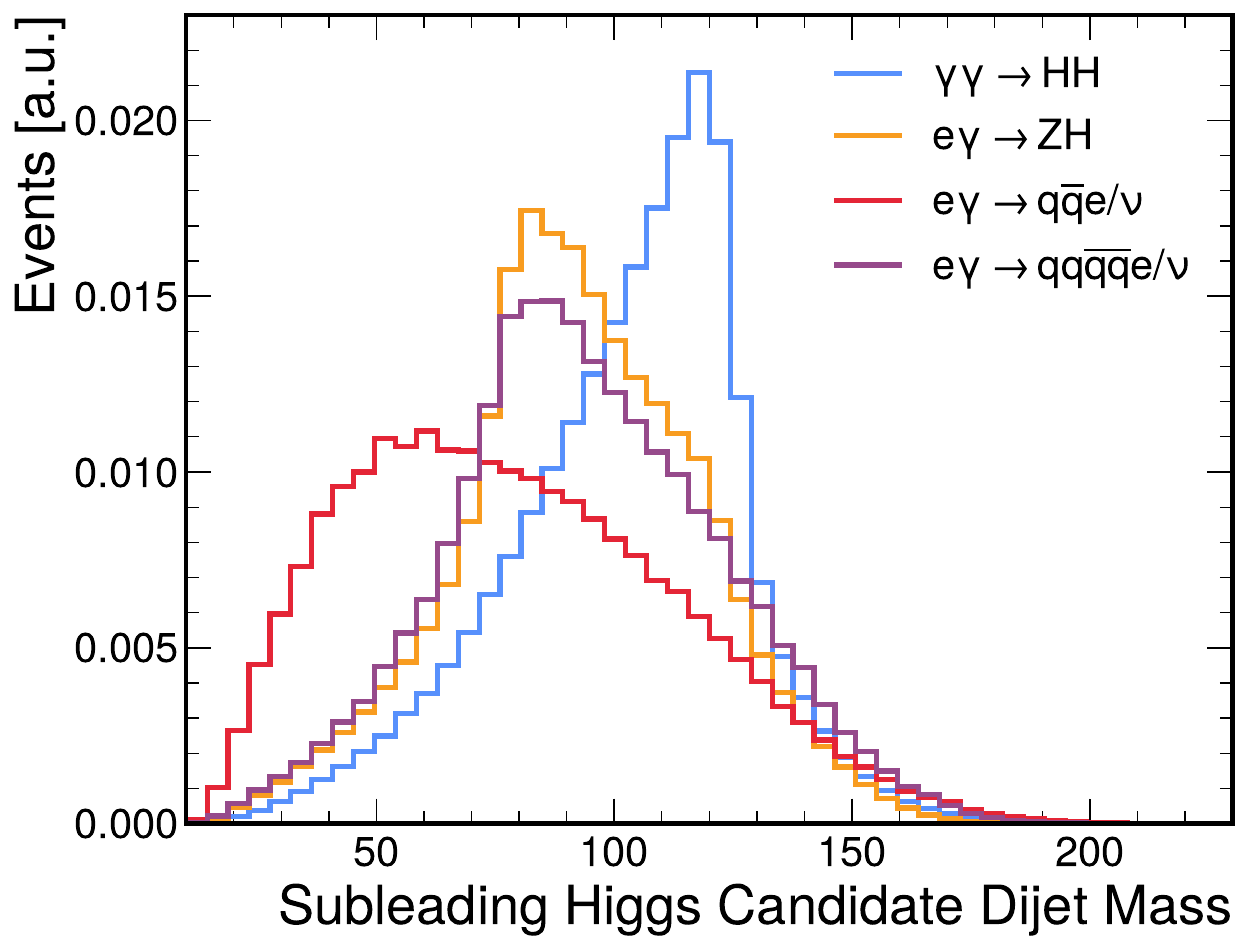}
    \end{minipage}
    \caption{Invariant mass distributions of the leading (left) and subleading (right) Higgs dijet candidates for the signal and the $\gamma\gamma$ (top), $e^+e^-$ (middle), and $e\gamma$ (bottom) backgrounds. The distributions are normalized such that area under the curve is unity.}
    \label{fig:Higgs_mass_dists_AA}
\end{figure}

PFOs are then clustered into jets using the exclusive Durham $k_T$ algorithm~\cite{GavinDurham,CataniClustering,MorettiClustering}, widely used in $e^+e^-$ colliders, which combines particles iteratively based on the distance metric:
\begin{equation}
\label{eq:yij}
    d_{ij} = 2 \min(E_i^2, E_j^2) (1 - \cos \theta_{ij}),
\end{equation}
where $\theta_{ij}$ is the angle between particles $i$ and $j$. At each step, the pair with the smallest $d_{ij}$ is combined into a new pseudo-particle using the $E$-scheme (four-momentum summation), and the process repeats until either a predefined number of jets, $n$, is reached. All events are clustered into four jets using the exclusive $n$-jets mode of Durham with $n=4$. After preselection, Higgs boson candidates are formed by choosing the jet-pair combination with the smallest $\chi^2$, defined as:

\begin{equation}
\label{eq:chi2}
\chi^2 = \frac{(m_{ij} - m_H)^2}{\sigma_m^2} + \frac{(m_{kl} - m_H)^2}{\sigma_m^2},
\end{equation}

where $\sigma_m$ represents the reconstructed Higgs mass jet-pair resolution, $m_H=$125~GeV is the Higgs boson mass, and $m_{ij}$ and $m_{kl}$ are the invariant masses of any combination of jet pairs. After the jet pairs are formed, they are labeled as leading and subleading, where the leading jet pair contains the leading jet (i.e.~the jet with the highest $E$). Fig.~\ref{fig:Higgs_mass_dists_AA} shows the invariant mass distributions of the Higgs dijet candidates for the signal and dominant backgrounds. Our analysis strategy makes use of these distributions as inputs for our machine learning algorithm, in addition to other observables, as described in later sections.

\section{Analysis Strategy}
\label{analysis}

The initial event preselection is designed to isolate signal-like topologies while suppressing the large backgrounds present in $\gamma\gamma$ collisions. As the signal process $\gamma\gamma \rightarrow HH \rightarrow bb\overline{bb}$ produces a fully hadronic final state with exactly four $b$-jets, four Durham jets were required, at least three b-tagged, and no isolated leptons. Table \ref{tab:eventsPreselection} updates Table \ref{tab:backgrounds} with the number of events after the preselection. 

\begin{table}[htbp]
    \centering
    \begin{tabular}{lll}
        \hline
        \textbf{Process} & \textbf{Initial Events} & \textbf{After Preselection} \\
        \hline
        \boldmath$\gamma\gamma \rightarrow HH \rightarrow bb\overline{bb}$\unboldmath      & \textbf{1,812}      & \textbf{703} \\
        \hline
        $\gamma\gamma \rightarrow W^+W^-\rightarrow 
        q\overline{q}q\overline{q}$           & 3,813,000      &  4,717  \\
        $\gamma\gamma \rightarrow t\overline{t}$         & 2,866,000     & 34,695 \\
        $\gamma\gamma \rightarrow ZZ\rightarrow q\overline{q}q\overline{q}$               & 1,378,000    & 98,527 \\
        $\gamma\gamma \rightarrow q\overline{q}$         & 307,700     &  11,112 \\
        $\gamma\gamma \rightarrow ZH\rightarrow q\overline{q}H$               & 8,202      &1,004 \\
        \hline
        $e\gamma \rightarrow eq\overline{q}\textrm{ or }\nu q\overline{q}$              & 41,195      & 1,206 \\
        $e\gamma \rightarrow eq\overline{q}q\overline{q}\textrm{ or }\nu q\overline{q}q\overline{q}$       & 7,681     & 137  \\
        $e\gamma \rightarrow eq\overline{q}H$             & 3,282     & 217 \\
        \hline
        $e^+e^- \rightarrow q\overline{q}$               & 753,615     & 54,067 \\
        $e^+e^- \rightarrow W^+W^- \textrm{ or } ZZ\rightarrow q\overline{q}q\overline{q}$        & 152,212     &  14,121 \\
        $e^+e^- \rightarrow ZH \rightarrow q\overline{q}H$                     & 123,698      &  13,771 \\
        $e^+e^- \rightarrow t\overline{t}$               & 57,001     & 648 \\
        \hline
    \end{tabular}
    \caption{Number of events after preselection for signal and backgrounds.}
    \label{tab:eventsPreselection}
\end{table}

Twelve boosted decision tree \cite{10.1214/aos/1013203451} (BDT) models were trained in order to separate signal from each of the backgrounds. Specifically, for each background process, a BDT is trained to discriminate between that particular background and the signal, resulting in one binary classifier per background. The canonical XGBoost \cite{Chen_2016} library is employed as the BDT implementation, a gradient boosting framework that sequentially trains decision trees to minimize classification loss. Each model is trained using binary cross-entropy loss to distinguish signal (label 1) from background (label 0). The XGBoost hyperparameters were selected through manual tuning and set as follows: maximum tree depth of 4, learning rate of 0.05, 500 boosting rounds with early stopping after 50 rounds of no validation improvement, subsample ratio of 0.8, and column subsample ratio of 0.8 for tree construction. L2 regularization (with $\lambda = 1.0$) is also applied to prevent overfitting. Data are split into training, validation, and test set subsets using a 70-10-20 stratified split.  Model outputs are probability scores in [0, 1] representing the likelihood of an event being signal. 

The BDT models are trained on a comprehensive set of kinematic and substructure observable-based features extracted from 4-jet events. Individual jet kinematics are included for all four leading jets, comprising transverse momentum, energy ($E$), ($p_{\mathrm{T}}$), cosine of the polar angle ($\cos \theta$), azimuthal angle ($\phi$), and mass ($m$) for each jet. Pairwise combinations of jet features including: (i) the angular separation quantified by $\Delta R = \sqrt{\Delta \phi^2 + \Delta \eta^2}$, (ii) the dijet invariant mass, and (iii) the Durham clustering distance are also included. For triplet combinations of jets, the invariant mass of each triple is computed. The invariant mass of all four jets combined is also included. To enhance discrimination for Higgs boson resonances, the two Higgs candidate masses are extracted by minimizing a $\chi^2$ metric based on Eq.~\ref{eq:chi2} that penalizes deviations from the Higgs mass ($m_H = 125$ GeV) for each possible pairing of jets. The optimal pairing yields two Higgs candidates with their corresponding masses, transverse momenta, and angular coordinates. Finally, missing transverse energy (MET) kinematics, consisting of $p_{\mathrm{T}}^{\text{miss}}$, $\eta_{\text{miss}}$, and $\phi_{\text{miss}}$ are included to capture information about undetected particles. This feature set totals approximately 50 observables per event, enabling the BDT to learn complex decision boundaries that leverage both individual particle properties and the geometric structure of multi-particle final states.

In order to combine each of the models into a single optimal discriminant between signal and all backgrounds, a Genetic Algorithm \cite{10.5555/534133} (GA), which uses the outputs from the BDTs as its 12 input features, was used. The GA determines, for each of the 12 distributions, the optimal selection thresholds that together maximize signal significance, defined as $\sigma = S/{\sqrt{S + B}}$, where $S$ is the number of remaining signal events and $B$ is the number of background events that survive for a given cut. The genetic algorithm uses a population size of 120. Tournament selection is then run with size 5 to identify optimal parents. The next generation is generated through blend crossover with blend factor $\alpha = 0.5$, with probability 0.7, and Gaussian mutation with standard deviation 0.3, with probability 0.3 per individual and 0.3 per coordinate. The algorithm runs for 50 generations, maintaining a Hall of Fame to track the best threshold configuration encountered. Upon convergence, the optimal threshold vector is applied to all datasets to determine final signal and background event survival rates. The Distributed Evolutionary Algorithms in Python (\textsc{DEAP}) \cite{DEAP_JMLR2012} library is used to implement the GA framework.

\begin{figure} 
    \centering
    \begin{minipage}{0.32\linewidth}
    \centering   \includegraphics[width=\linewidth]{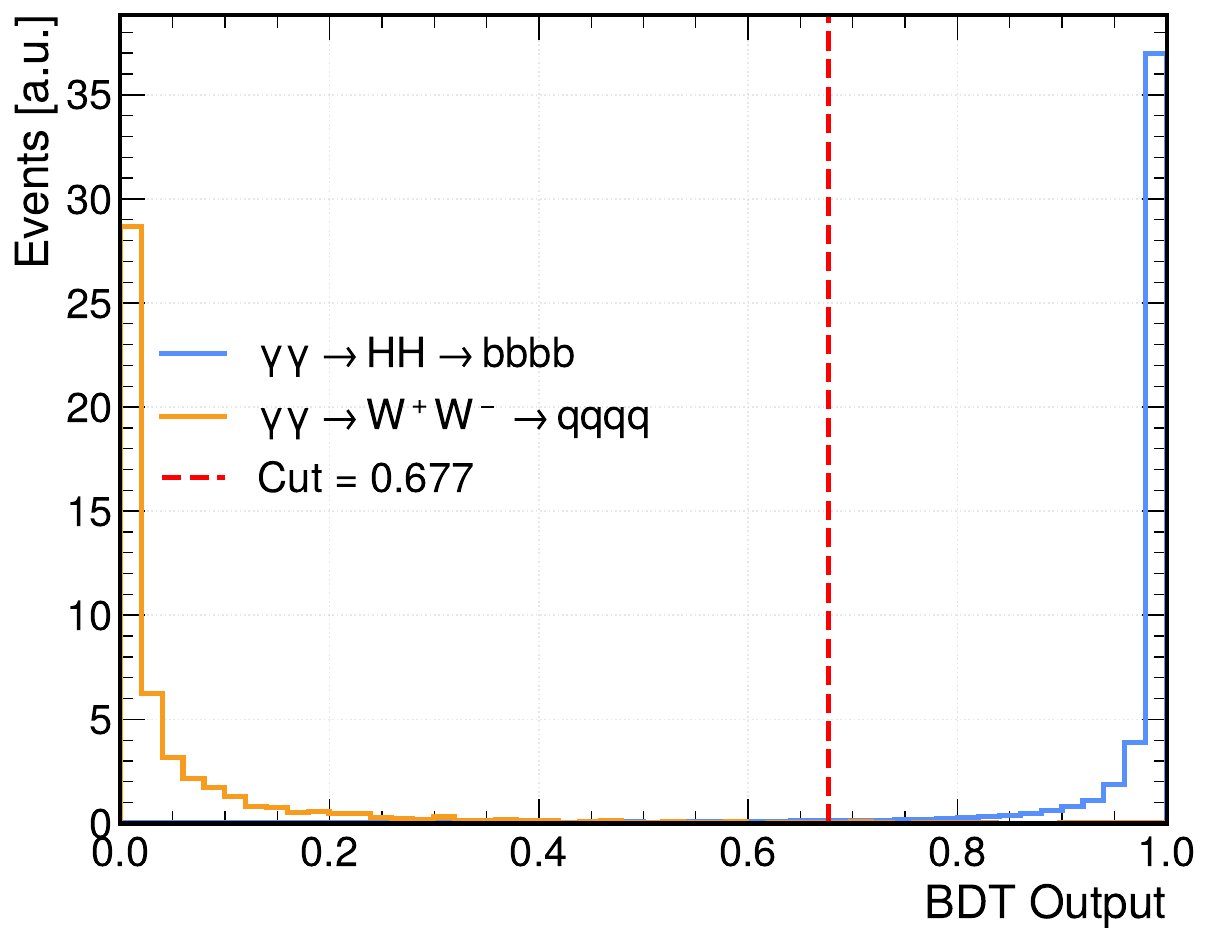}
    \end{minipage}
    \begin{minipage}{0.32\linewidth}
    \centering   \includegraphics[width=\linewidth]{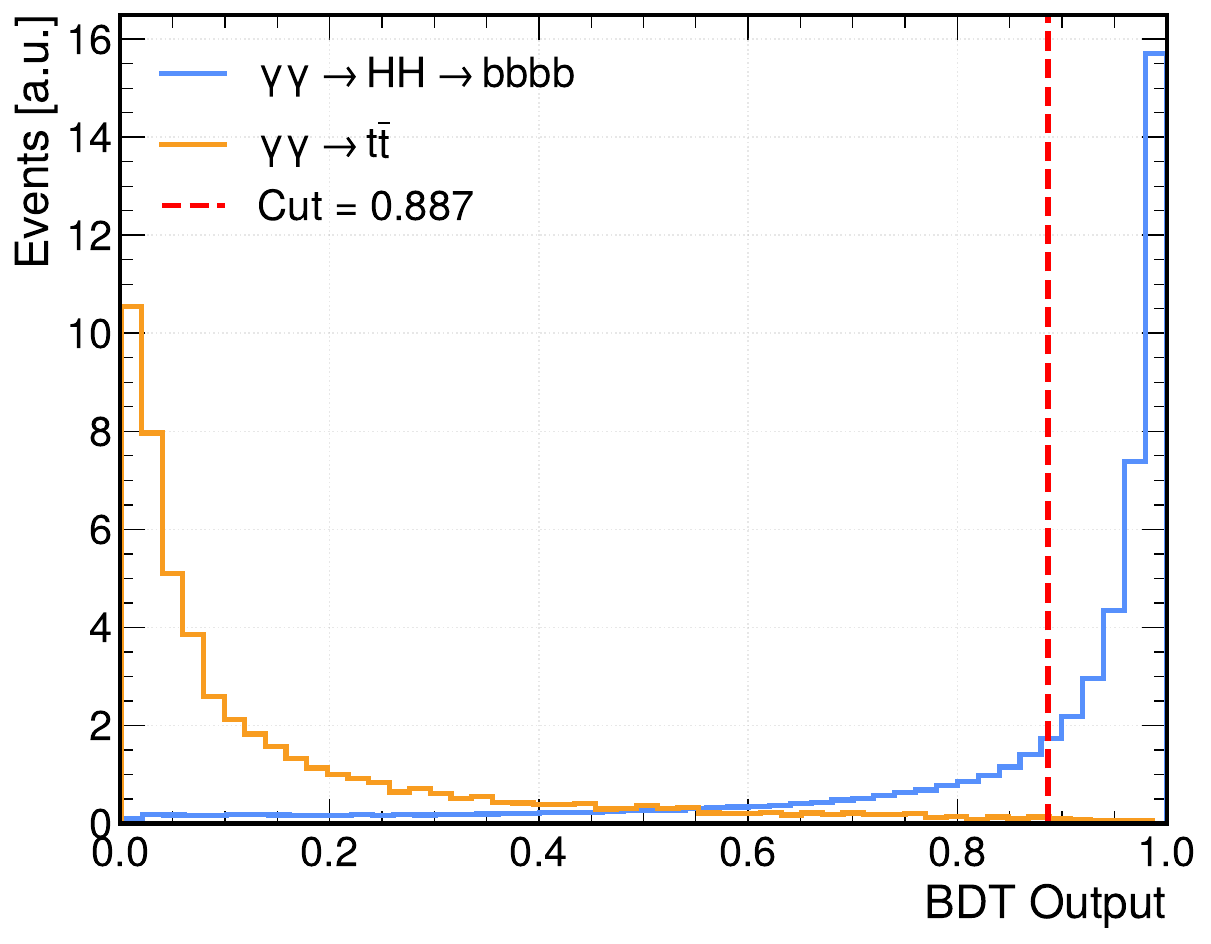}
    \end{minipage}
    \begin{minipage}{0.32\linewidth}
    \centering   \includegraphics[width=\linewidth]{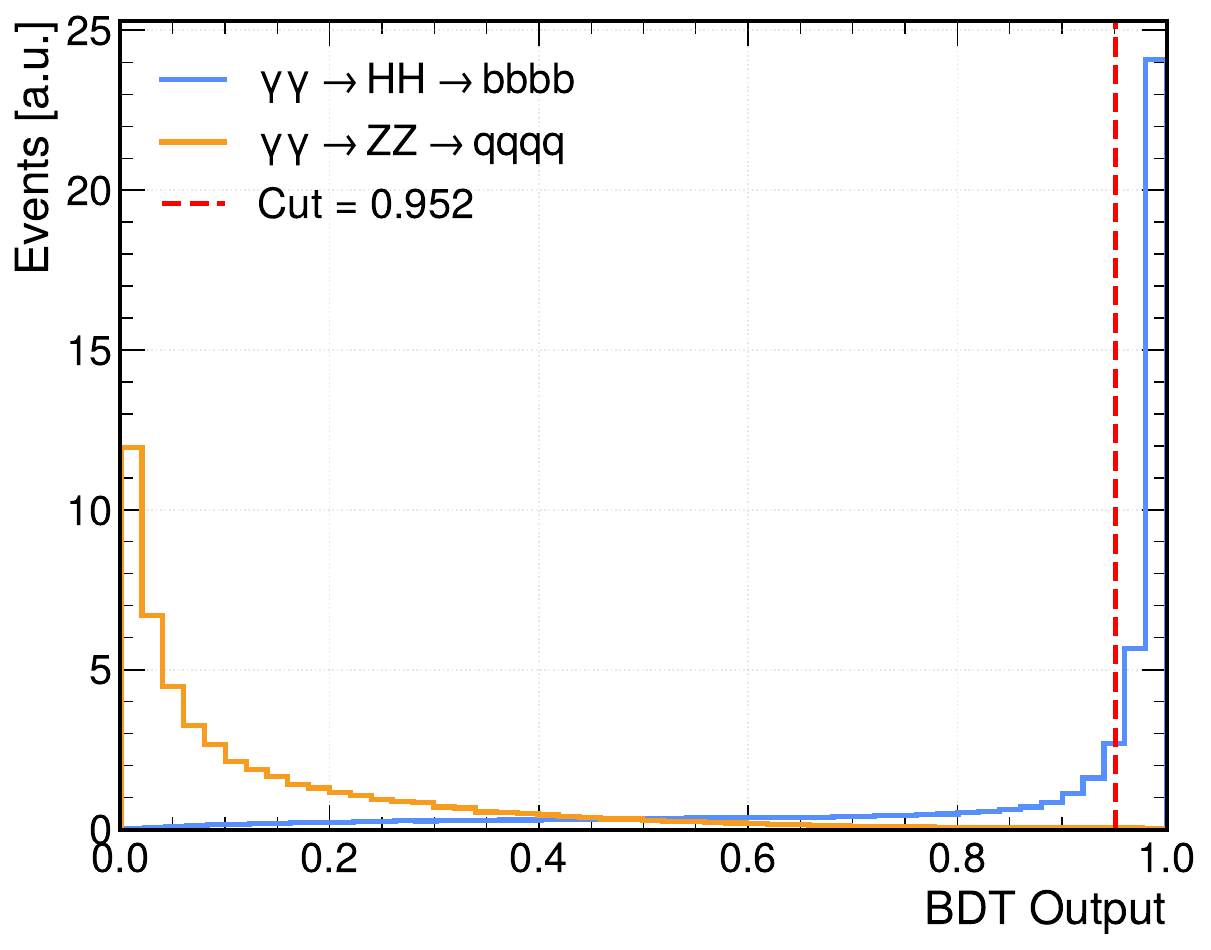}
    \end{minipage}

    \begin{minipage}{0.32\linewidth}
    \centering   \includegraphics[width=\linewidth]{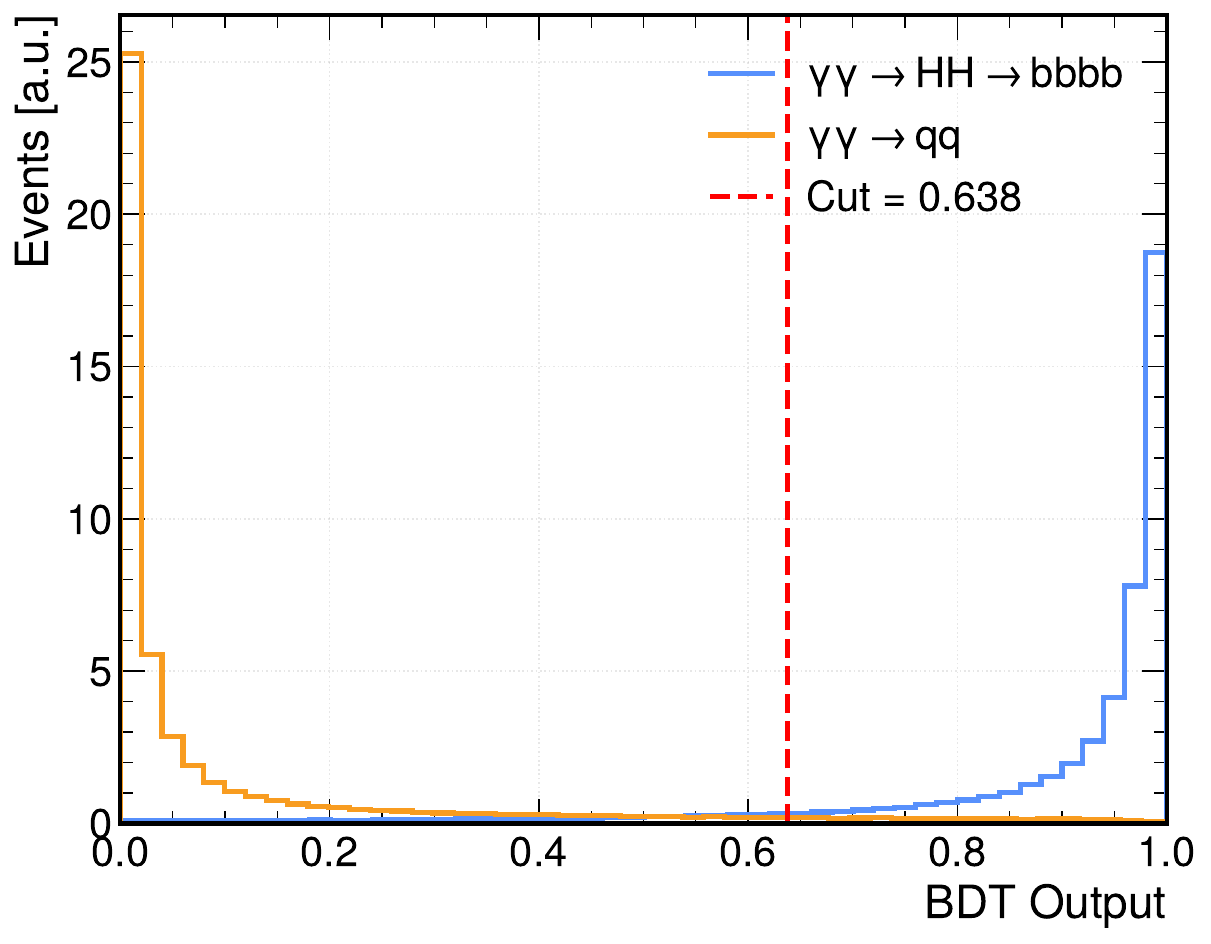}
    \end{minipage}
    \begin{minipage}{0.32\linewidth}
    \centering   \includegraphics[width=\linewidth]{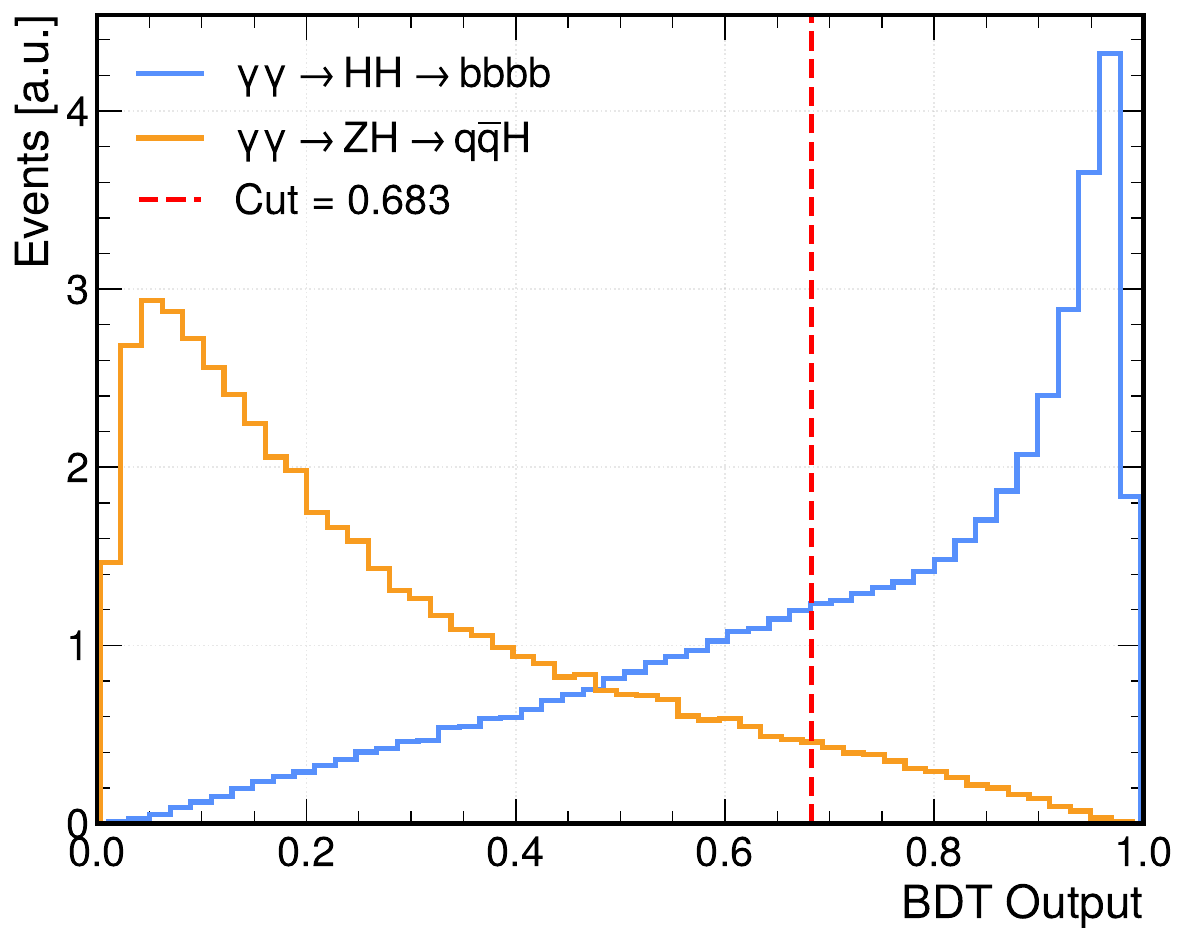}
    \end{minipage}
    \begin{minipage}{0.32\linewidth}
    \centering   \includegraphics[width=\linewidth]{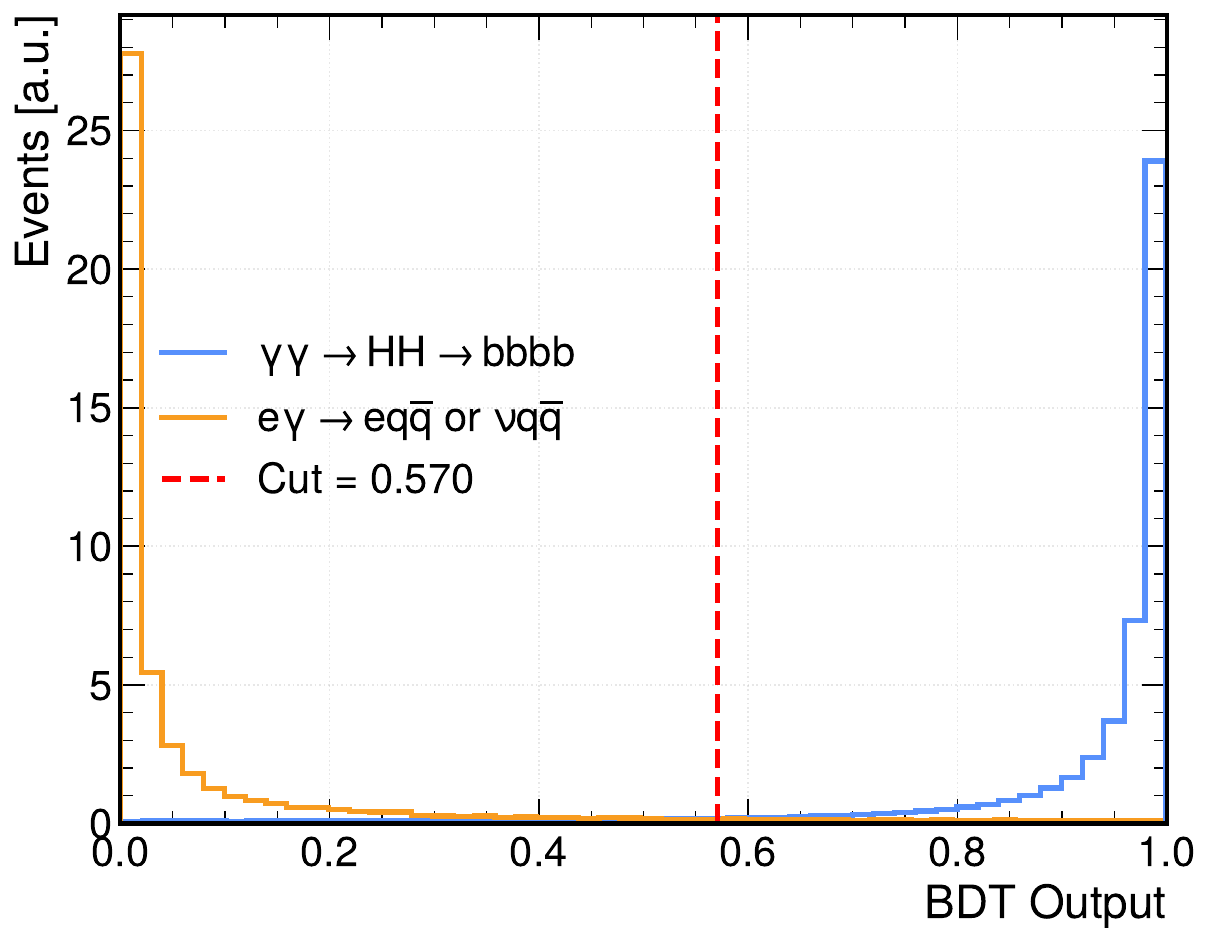}
    \end{minipage}

    \begin{minipage}{0.32\linewidth}
    \centering   \includegraphics[width=\linewidth]{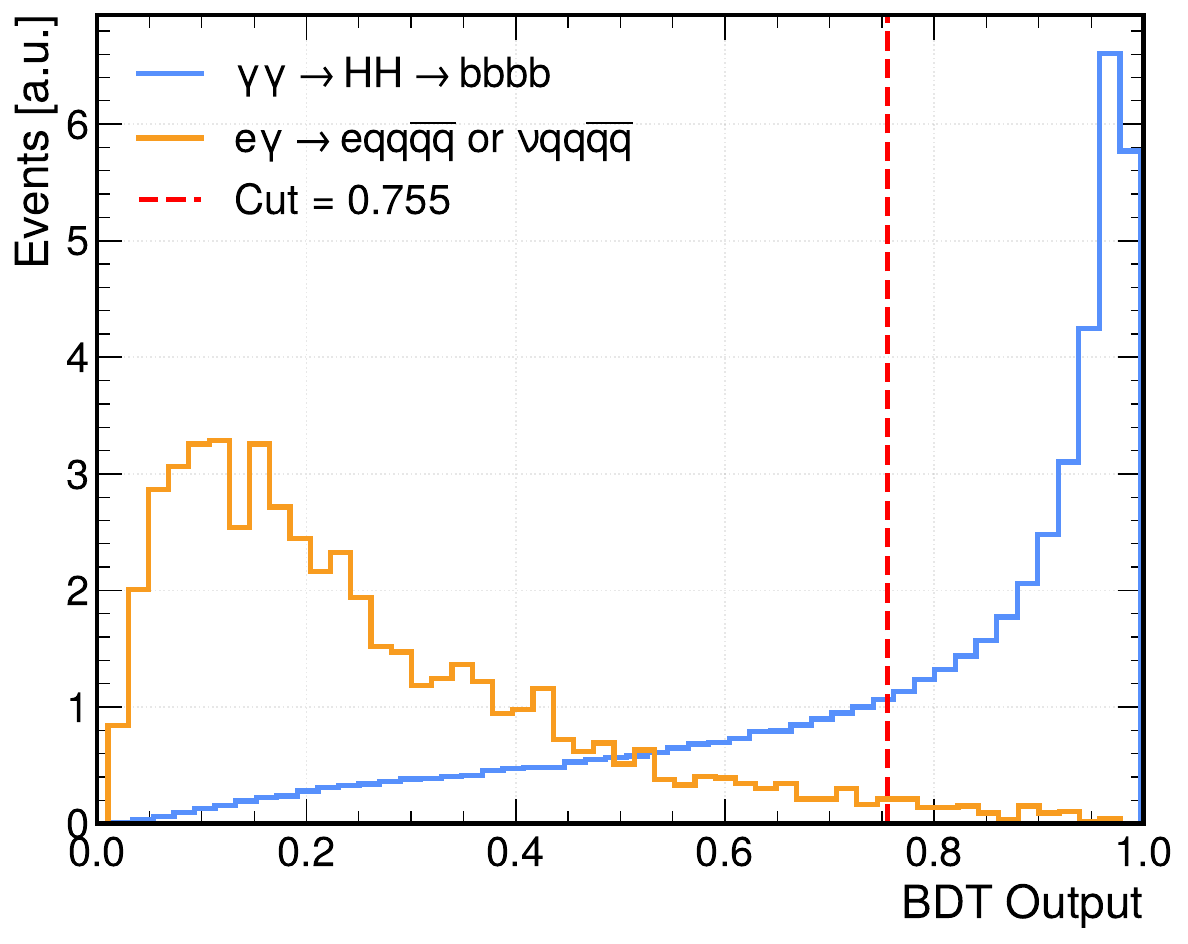}
    \end{minipage}
    \begin{minipage}{0.32\linewidth}
    \centering   \includegraphics[width=\linewidth]{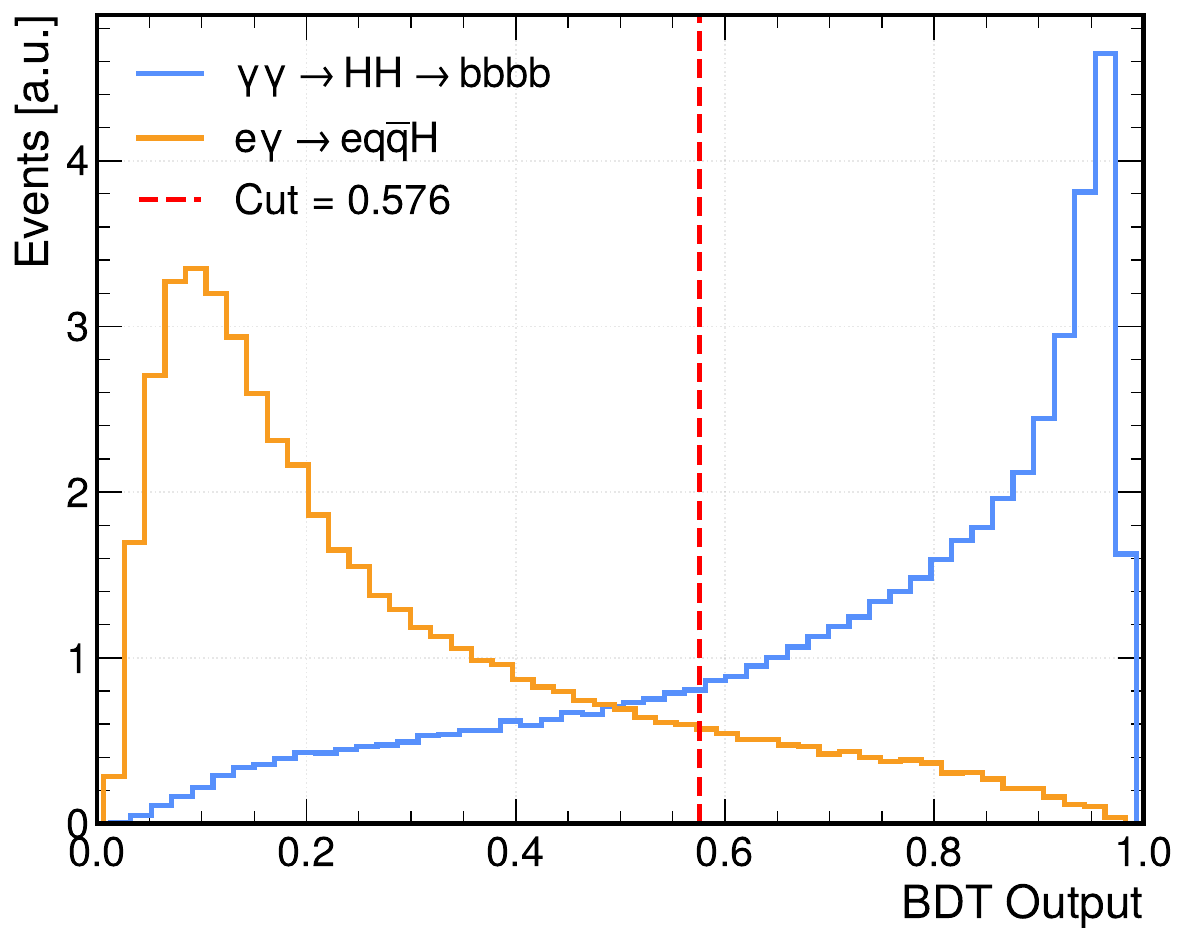}
    \end{minipage}
    \begin{minipage}{0.32\linewidth}
    \centering   \includegraphics[width=\linewidth]{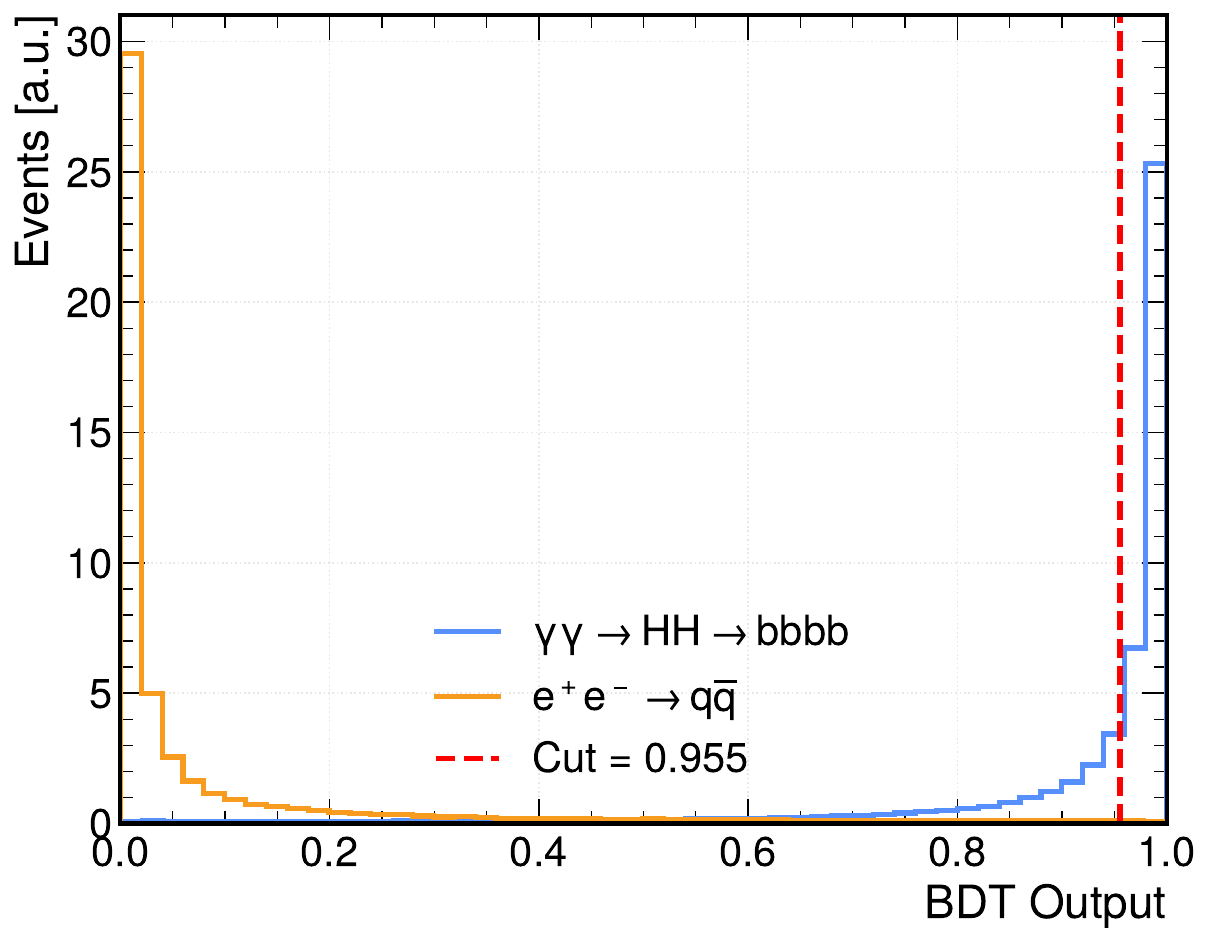}
    \end{minipage}

    \begin{minipage}{0.32\linewidth}
    \centering   \includegraphics[width=\linewidth]{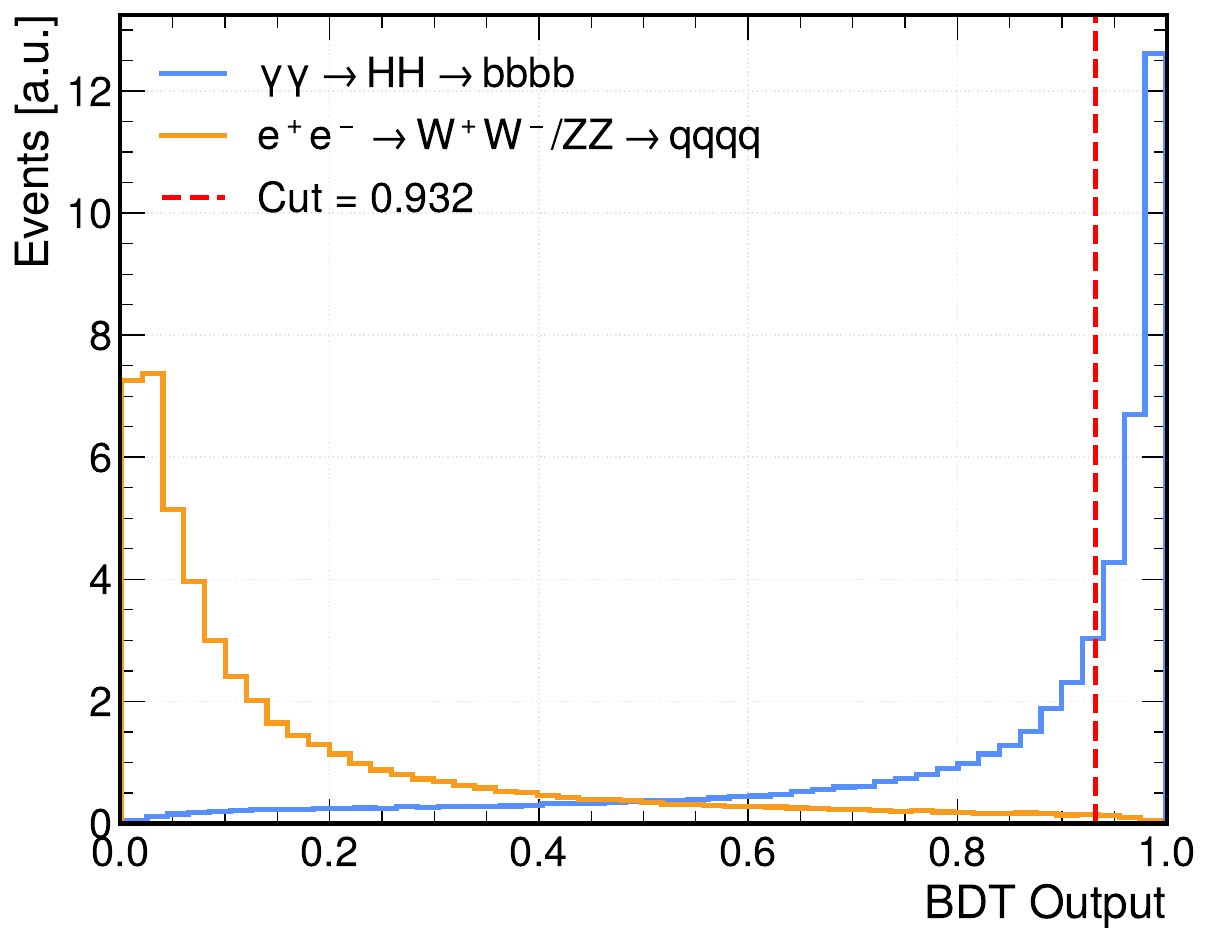}
    \end{minipage}
    \begin{minipage}{0.32\linewidth}
    \centering   \includegraphics[width=\linewidth]{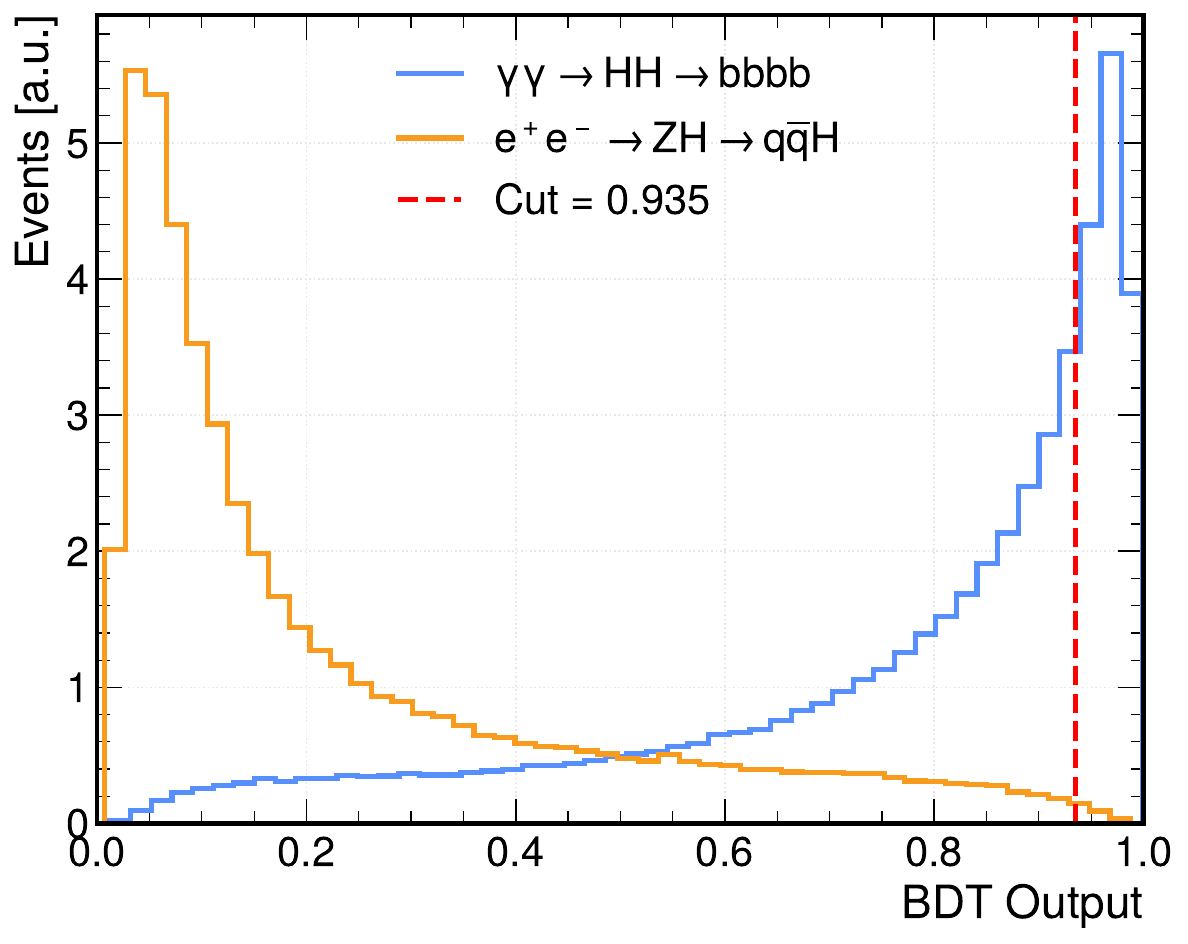}
    \end{minipage}
    \begin{minipage}{0.32\linewidth}
    \centering   \includegraphics[width=\linewidth]{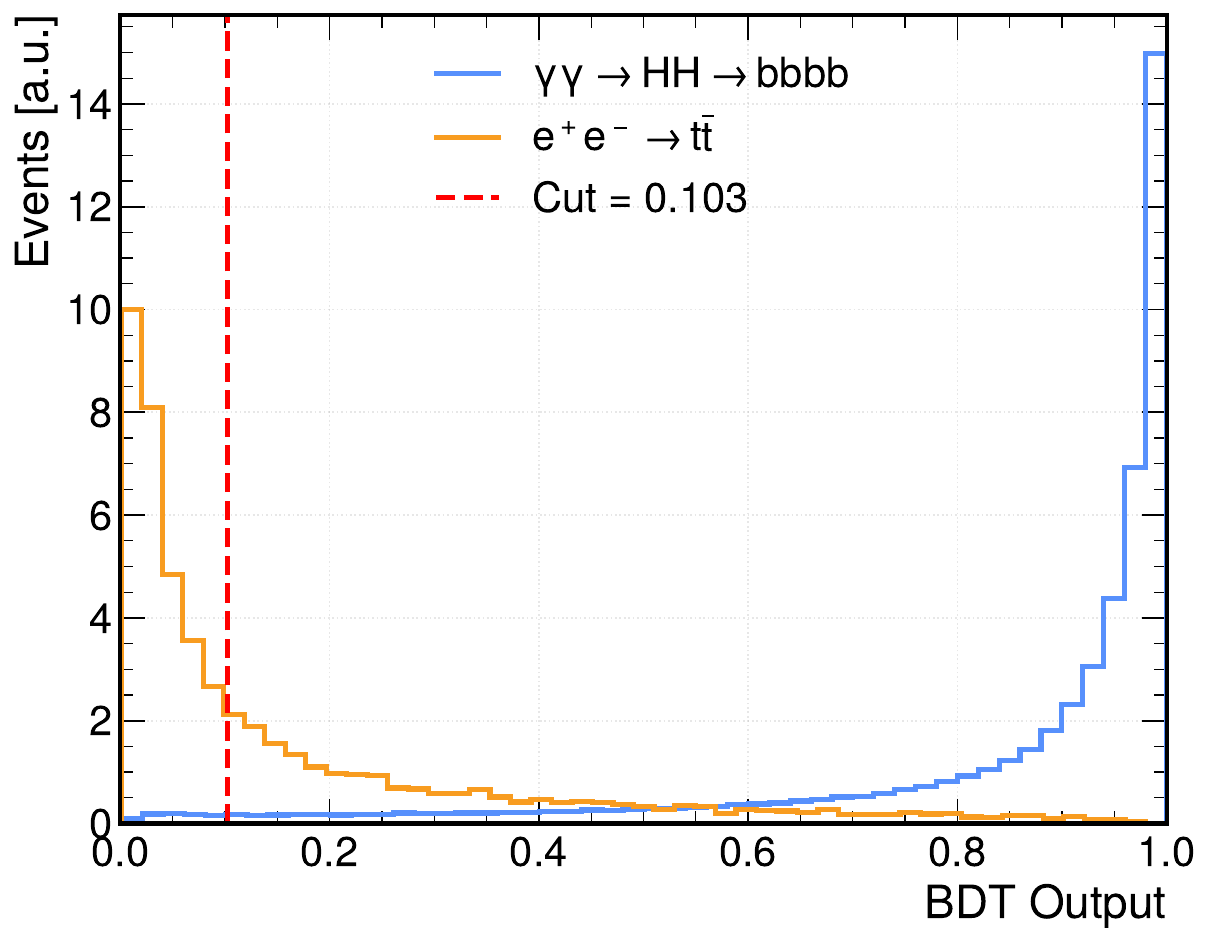}
    \end{minipage}
    \caption{BDT output scores for each signal vs. background pair. Each model receives the name of the background it was trained with. The cut found by the GA for each BDT is marked with a red dashed line. (1,1): $\mathrm{BDT}_{\gamma \gamma \rightarrow W^+W^- \rightarrow q\overline{q}q\overline{q}}$; (1,2): $\mathrm{BDT}_{\gamma \gamma \rightarrow t\overline{t}}$; (1,3): $\mathrm{BDT}_{\gamma \gamma \rightarrow ZZ \rightarrow q\overline{q}qq}$; (2,1): $\mathrm{BDT}_{\gamma \gamma \rightarrow q\overline{q}}$; (2,2): $\mathrm{BDT}_{\gamma \gamma \rightarrow ZH \rightarrow q\overline{q}H}$; (2,3): $\mathrm{BDT}_{e \gamma \rightarrow eq\overline{q}/\nu q\overline{q}}$; (3,1): $\mathrm{BDT}_{e \gamma \rightarrow eq\overline{q}q\overline{q}/\nu q\overline{q}q\overline{q}}$; (3,2): $\mathrm{BDT}_{e \gamma \rightarrow eq\overline{q}H}$; (3,3): $\mathrm{BDT}_{e^+e^- \rightarrow q\overline{q}}$; (4,1): $\mathrm{BDT}_{e^+e^- \rightarrow W^+W^-/ZZ \rightarrow q\overline{q}q\overline{q}}$; (4,2): $\mathrm{BDT}_{e^+e^- \rightarrow ZH \rightarrow q\overline{q}H}$; (4,3): $\mathrm{BDT}_{e^+e^- \rightarrow t\overline{t}}$.}
    \label{fig:outputXGB}
\end{figure}

Fig.~\ref{fig:outputXGB} shows the distribution of the BDT output scores for each signal versus background, noting the cut found by the GA.

\section{Results}
\label{sec:results}

\begin{table}[htbp]
    \centering
    \begin{tabular}{llll}
        \hline
        \textbf{Process} & \textbf{Initial Events} & \textbf{After Preselection} &  \textbf{After GA} \\
        \hline
        \boldmath$\gamma\gamma \rightarrow HH \rightarrow bb\overline{bb}$\unboldmath      & \textbf{1,812}      & \textbf{703} & \textbf{151} \\
        \hline
        $\gamma\gamma \rightarrow W^+W^-\rightarrow 
        q\overline{q}q\overline{q}$           & 3,813,000      &  4,717 & 2  \\
        $\gamma\gamma \rightarrow t\overline{t}$         & 2,866,000     & 34,695 & 29 \\
        $\gamma\gamma \rightarrow ZZ\rightarrow q\overline{q}q\overline{q}$               & 1,378,000    & 98,527 & 8 \\
        $\gamma\gamma \rightarrow q\overline{q}$         & 307,700     &  11,112 & 6 \\
        $\gamma\gamma \rightarrow ZH\rightarrow q\overline{q}H$               & 8,202      &1,004 & 9\\
        \hline
        $e\gamma \rightarrow eq\overline{q}\textrm{ or }\nu q\overline{q}$              & 41,195      & 1,206 & 1 \\
        $e\gamma \rightarrow eq\overline{q}q\overline{q}\textrm{ or }\nu q\overline{q}q\overline{q}$       & 7,681     & 137  & 0\\
        $e\gamma \rightarrow eq\overline{q}H$             & 3,282     & 217 & 0 \\
        \hline
        $e^+e^- \rightarrow q\overline{q}$               & 753,615     & 54,067 & 12\\
        $e^+e^- \rightarrow W^+W^- \textrm{ or } ZZ\rightarrow q\overline{q}q\overline{q}$        & 152,212     &  14,121 &2 \\
        $e^+e^- \rightarrow ZH \rightarrow q\overline{q}H$                     & 123,698      &  13,771 & 62 \\
        $e^+e^- \rightarrow t\overline{t}$               & 57,001     & 648 & 1 \\
        \hline
    \end{tabular}
    \caption{Number of events after GA cuts on the BDT output distributions for the signal and backgrounds.}
    \label{tab:eventsPreselection_GA}
\end{table}

\begin{figure}
    \centering   
    \begin{minipage}{0.495\linewidth}
    \centering
        \includegraphics[width=\linewidth]{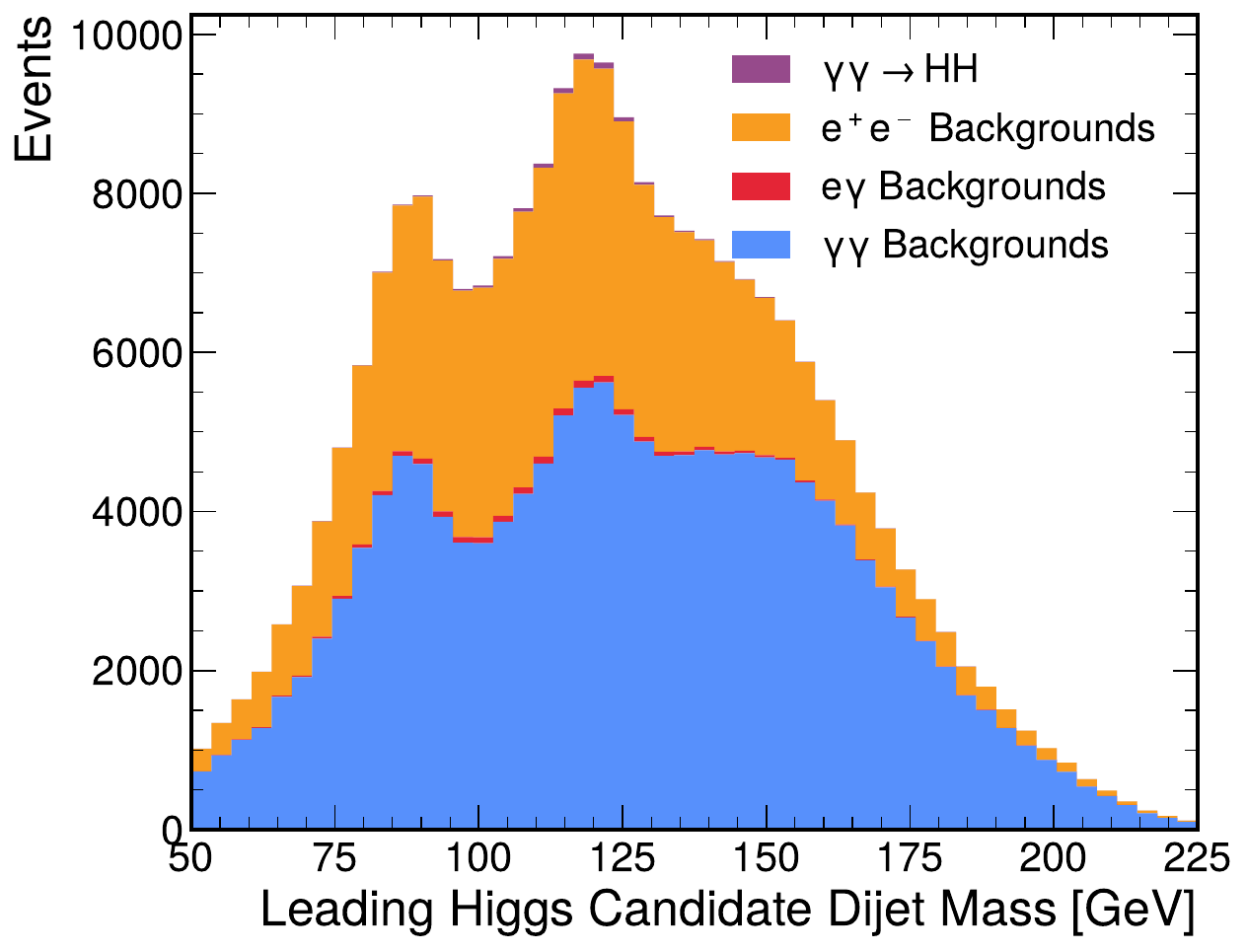}
    \end{minipage}
    \begin{minipage}{0.495\linewidth}
    \centering
        \includegraphics[width=\linewidth]{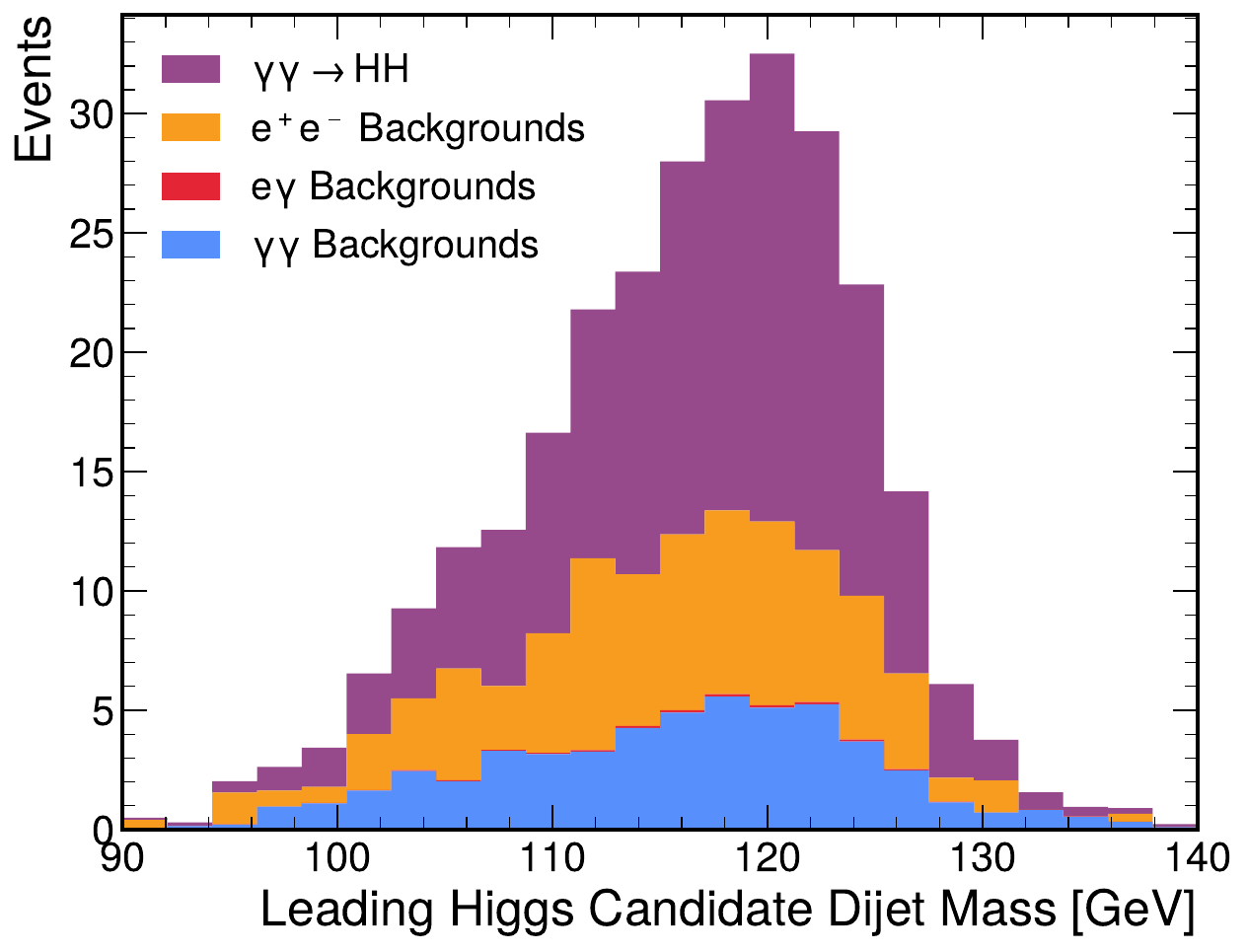}
    \end{minipage}
    \caption{Invariant mass distributions of the leading Higgs dijet candidate for the $\gamma\gamma \to HH$ signal (purple), $\gamma\gamma$ backgrounds (blue), $e\gamma$ backgrounds (red), and $e^+e^-$ backgrounds (orange) before (left) and after (right) the GA cuts on the BDT output distributions.}
    \label{fig:Higgs_mass_dists_before_after_GA}
\end{figure}
\begin{figure}[htbp]
    \centering
    \includegraphics[width=0.7\textwidth]{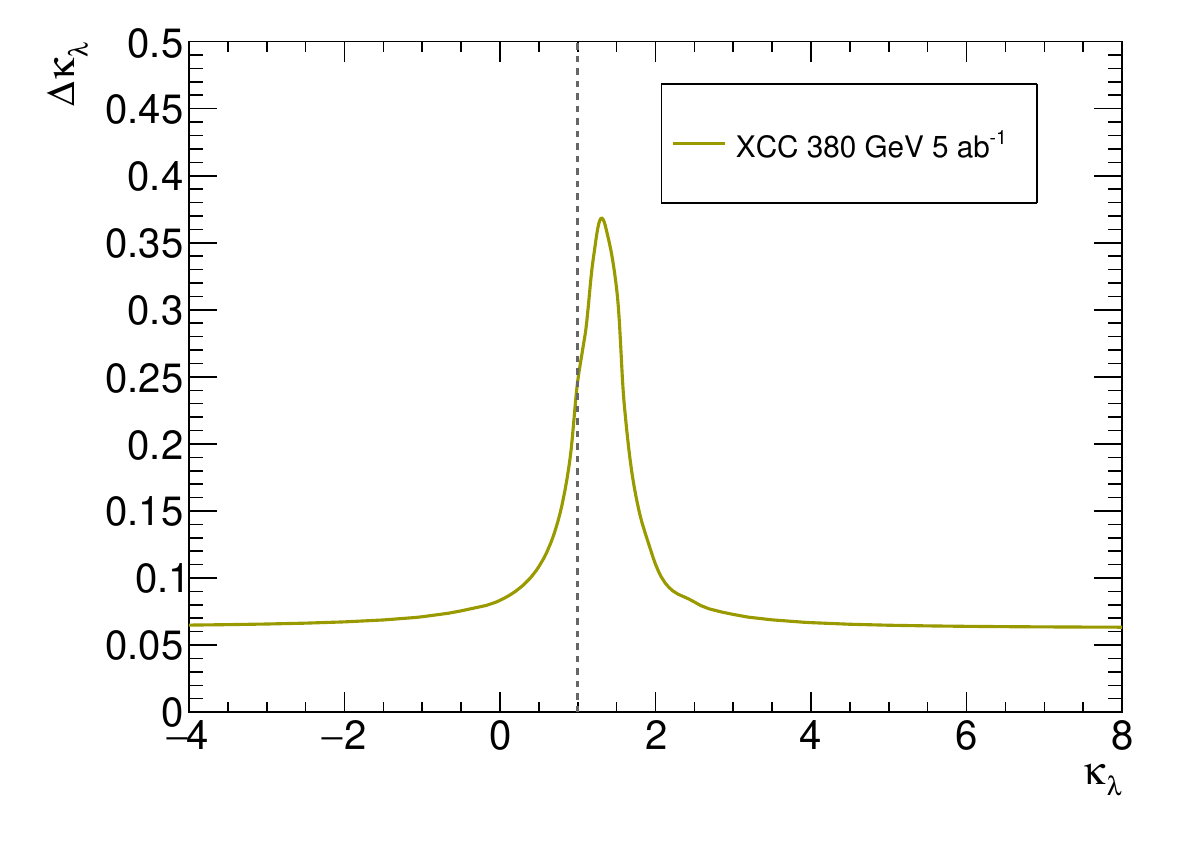}
    \caption{$\Delta\kappa_\lambda$ versus $\kappa_\lambda$ derived from an 11.1\% $\gamma\gamma\rightarrow HH$ cross-section measurement error at $\sqrt{s} = 380$~GeV with 4.9~ab$^{-1}$ luminosity for $260\ \mathrm{GeV} < \sqrt{\widehat{s}} < 380 \ \mathrm{GeV}$ and $\kappa_\lambda=1$.  The 11.1\% error corresponds to the analysis of $\gamma \gamma \rightarrow HH \rightarrow bb\overline{bb}$ described in this paper. }
    \label{fig:delkLam_xcc380_bbbb}
\end{figure}

Table \ref{tab:eventsPreselection_GA} and Fig.~\ref{fig:Higgs_mass_dists_before_after_GA} show the number of events, and the leading Higgs mass distribution, before and after the multivariate discriminant. The statistical significance of the $\gamma\gamma \rightarrow HH \rightarrow bb\overline{bb}$ channel is estimated as $\sigma = S/\sqrt{S+B} = 8.97$, corresponding to about a 11.1\% error on the measured value of $\sigma_{HH}$
at $\sqrt{s}=380$, which can be converted into an
error on $\kappa_\lambda$, the Higgs self-coupling relative to its SM value. For this, we use the gold $\sigma$ versus $\kappa_\lambda$ curve in Fig.~\ref{fig:sigHH_vs_kLam} under the assumption that the error scales with the square root of the number of signal events:
\begin{equation}
\frac{\Delta\sigma}{\sigma}=\frac{\Delta \sigma_0} {\sigma_0}\sqrt{\frac{\sigma_0}{\sigma}}=0.111\sqrt{\frac{\sigma_0}{\sigma}} \ ,
\end{equation}
where $\sigma_0$ is the $\gamma\gamma\rightarrow HH$ cross-section at $\kappa_\lambda=1$. For regions away from the minimum at $\kappa_\lambda=1.3$ the error on
$\kappa_\lambda$ is simply 
\begin{equation} \label{eq:delkapderiv}
\Delta\kappa_\lambda=\left| \frac{d\sigma}{d\kappa_\Lambda}\right|^{-1}\Delta\sigma=0.111\left| \frac{d\sigma}{d\kappa_\lambda}\right|^{-1}\sqrt{\sigma\sigma_0}\ .
\end{equation}
For the region near the minimum $\kappa_\lambda\approx1.3$ the error
$\Delta\kappa_\lambda$ is calculated by viewing the gold curve in Fig.~\ref{fig:sigHH_vs_kLam} as a double-valued $\kappa_\lambda$ versus cross-section plot
and solving for $\kappa_\lambda(\sigma\pm \Delta\sigma)$:
\begin{equation}
\Delta\kappa_\lambda=\frac{1}{2}\left[\kappa_\lambda(\sigma+\Delta\sigma)+\kappa_\lambda(\sigma-\Delta\sigma)\right]-\kappa_\lambda(\sigma)\ .
\end{equation}
The resulting values of $\Delta\kappa_\lambda$ versus $\kappa_\lambda$ are shown in Figure~\ref{fig:delkLam_xcc380_bbbb} assuming 4.9~ab$^{-1}$ luminosity for $260\ \mathrm{GeV} < \sqrt{\widehat{s}} < 380 \ \mathrm{GeV}$.

\begin{figure}[htbp]
    \centering
    \includegraphics[width=0.7\textwidth]{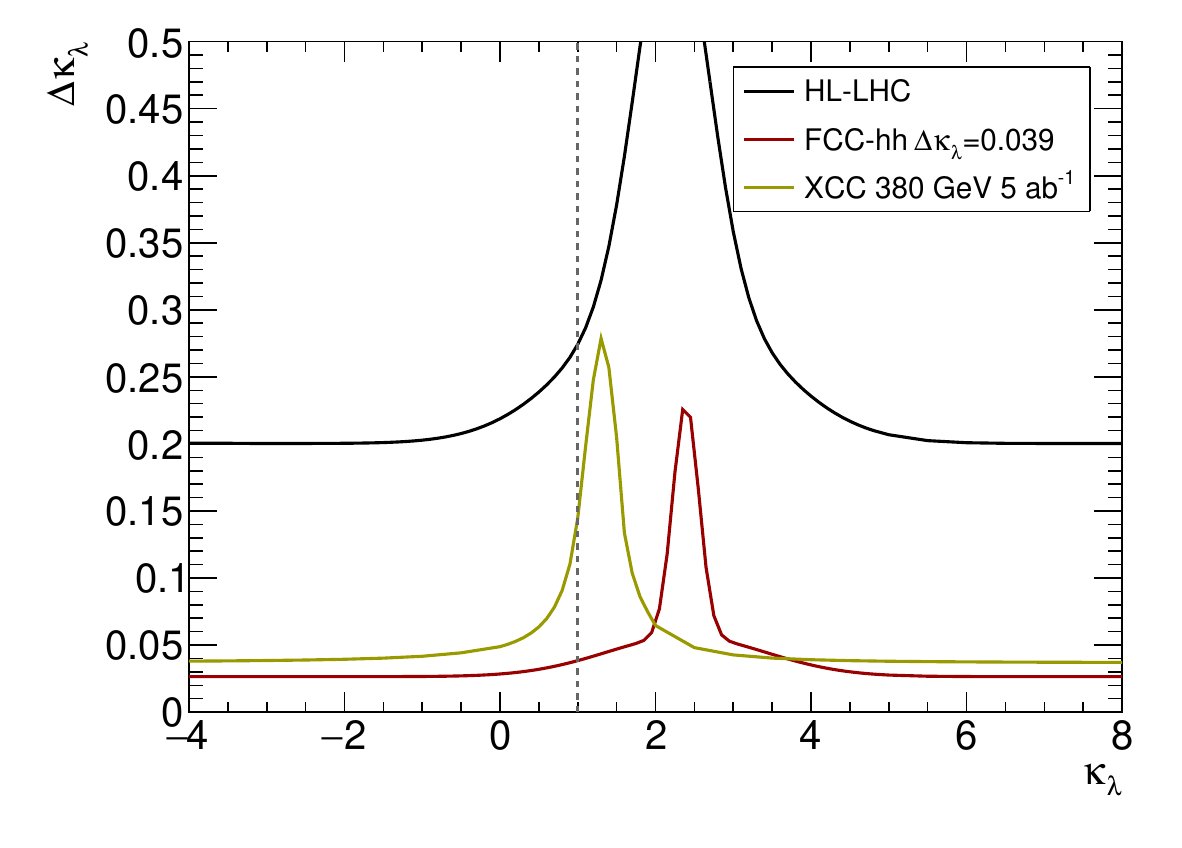}
    \caption{Prediction for the ultimate XCC $\Delta\kappa_\lambda$ measurement precision versus $\kappa_\lambda$ assuming a 6.5\% $\gamma\gamma\rightarrow HH$ cross-section measurement error at $\sqrt{s} = 380$~GeV with 4.9~ab$^{-1}$ luminosity for $260\ \mathrm{GeV} < \sqrt{\widehat{s}} < 380 \ \mathrm{GeV}$ and $\kappa_\lambda=1$ (gold curve). The 6.5\% error is a statistical extrapolation of the 11.1\% error for $\gamma\gamma\rightarrow HH \rightarrow bb\overline{bb} $ to a future analysis that would include all $HH$ decay topologies.  For comparison, the triple Higgs coupling sensitivity 
    of HL-LHC (black curve) and FCC-hh (red curve) are also shown.}
    \label{fig:delkLam_xcc380_HL-LHC_FCC-hh}
\end{figure}

Future analyses of $\gamma\gamma\rightarrow HH$ will utilize all  $HH$ events topologies, such as $b\overline{b}WW^*$, $b\overline{b}\gamma\gamma, c\overline{c}ZZ^*$, etc.
An estimate of $\Delta\sigma_{HH}=7\%\ =\ 11.1\%\times \sqrt{BR(H\rightarrow b\overline{b})^2}$ 
can be derived by  assuming that such an analysis has been performed and that all topologies produce the same $\gamma\gamma\rightarrow\ HH$ cross-section error as the $HH\rightarrow\ bb\overline{bb}$ topology, modulo statistics.
With an 
error of $\Delta\sigma_{HH}=7$\% for 
$\kappa_\lambda=1$ in hand, the prediction for $\Delta\kappa_\lambda$ versus $\kappa_\lambda$  at $\sqrt{s}=380$~GeV can be calculated using the gold curve in Figure~\ref{fig:sigHH_vs_kLam}; the result is displayed in Figure~\ref{fig:delkLam_xcc380_HL-LHC_FCC-hh}.  For comparison, the Higgs self-coupling sensitivity predictions for HL-LHC~\cite{ATLAS:2025eii} and FCC-hh \cite{Stapf:2023ndn,Baglio:2020ini} are also shown.


\begin{figure}[htbp]
    \centering
    \includegraphics[width=0.7\textwidth]{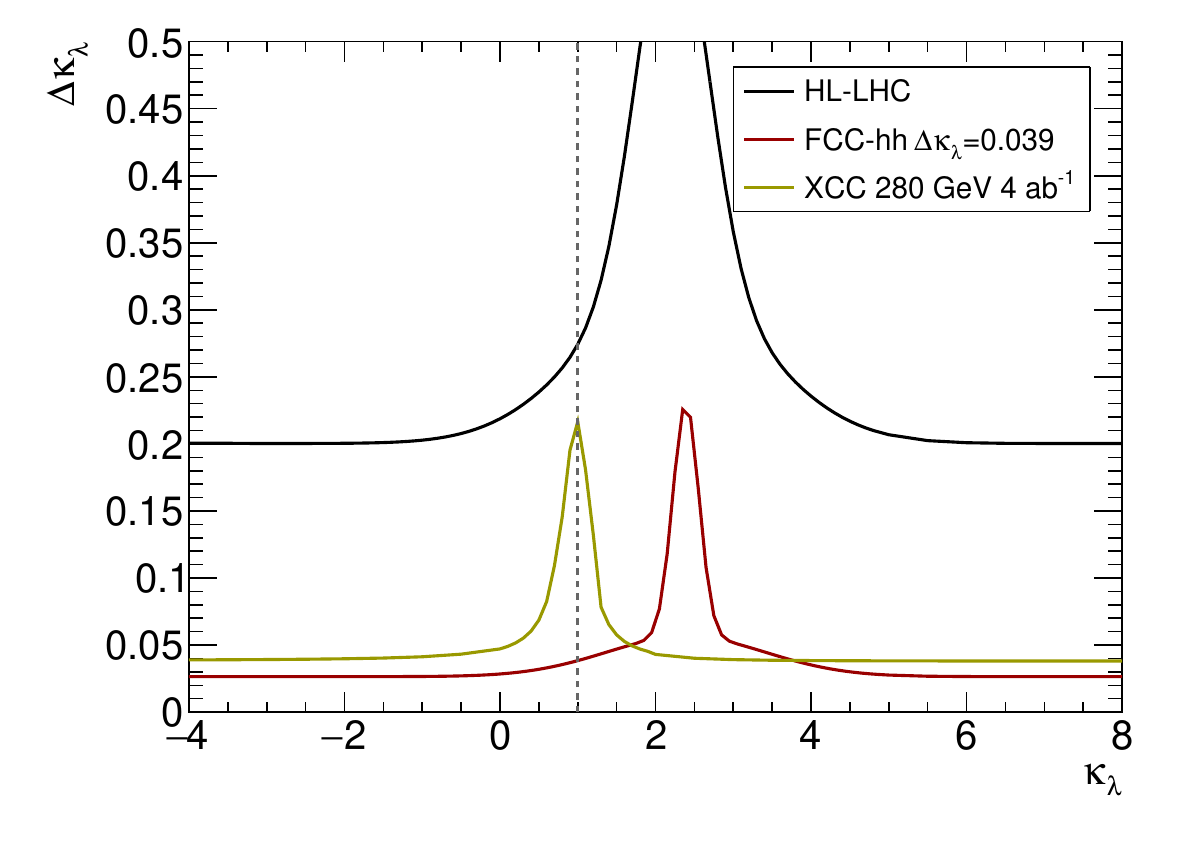}
    \caption{Prediction for the ultimate XCC $\Delta\kappa_\lambda$ measurement precision versus $\kappa_\lambda$ assuming an 11.8\% $\gamma\gamma\rightarrow HH$ cross-section measurement error at $\sqrt{s} = 280$~GeV with 4.9~ab$^{-1}$ luminosity for $250\ \mathrm{GeV} < \sqrt{\widehat{s}} < 280 \ \mathrm{GeV}$ and $\kappa_\lambda=1$ (gold curve). The 11.8\% error at $\sqrt{s} = 280$~GeV is an extrapolation of a 6.5\% error at $\sqrt{s} = 380$~GeV. For comparison, the triple Higgs coupling sensitivity 
    of HL-LHC (black curve) and FCC-hh (red curve) are also shown.}
    \label{fig:delkLam_xcc280_HL-LHC_FCC-hh}
\end{figure}

This analysis uses the center-of-mass energy of $\sqrt{s}=380$~GeV because it corresponds to the maximum $\gamma\gamma\rightarrow HH$ cross-section for $J_z=0$   and $\kappa_\lambda=1$.   
However, as pointed out in~\cite{Kawada:2012uy}, a $\gamma\gamma$ collider at $\sqrt{s}=280$~GeV should provide  better 
 Higgs self-coupling sensitivity for $\kappa_\lambda$ values different from 1 given the inverse relationship between $\Delta\kappa_\lambda$ and 
  $\left|d\sigma/d\kappa_\Lambda\right|$
 and the very large values of  $\left|d\sigma/d\kappa_\Lambda\right|$ for 
 $\sqrt{s}=280$~GeV (Figure~\ref{fig:sigHH_vs_kLam}).
The ultimate $\Delta\sigma_{HH}=6.5$\% error at $\sqrt{s}=380$~GeV assuming all double Higgs decay topologies have been analyzed can be extrapolated to a  11.8\% error at $\sqrt{s}=280$~GeV by scaling the $\sqrt{s}=380$~GeV  $\kappa_\gamma=1$ 
signal statistics to 280~GeV.  The prediction for $\Delta\kappa_\lambda$ versus $\kappa_\lambda$  at $\sqrt{s}=280$~GeV is then calculated using the magenta curve in Figure~\ref{fig:sigHH_vs_kLam}; the result is displayed in Figure~\ref{fig:delkLam_xcc280_HL-LHC_FCC-hh}.  

\newpage

\section{Conclusions}
\label{sec:conclusion}

In this paper, we investigated the potential of an XFEL-based $\gamma\gamma$ collider to measure the Higgs self-coupling. The use of high-energy X-ray lasers enables operation at large values of $x$, producing a sharply peaked luminosity spectrum that yields higher signal significance at substantially lower center-of-mass energies than previous $\gamma\gamma$ concepts based on optical lasers. In addition, the characteristic dependence of $\sigma(\gamma\gamma \to HH)$ on $\kappa_\lambda$ provides a complementary probe of the Higgs self-coupling relative to $e^+e^-$ and hadron colliders.

For a ten-year run at $\sqrt{s}=280$ GeV and $\kappa_\lambda = 1$, we estimate a measurement uncertainty of approximately $12\%$. For non-Standard-Model values of $\kappa_\lambda$, the projected uncertainty improves to 7–10\%, comparable to FCC-hh expectations. 

Our study includes all relevant $\gamma\gamma$, $e\gamma$, and $e^+e^-$ backgrounds arising from both the Compton and primary interaction regions and employs a fast simulation of an $e^+e^-$ linear-collider detector. However, several assumptions motivate further dedicated work. First, this study has excluded the region $0.95 \le \cos \theta \le 0.99$ due to the flux of soft X-rays from the Compton interaction regions. A follow-up study will implement a full detector simulation, where the impact of this background could be fully evaluated. These studies will also inform possible background suppression methods that could extend the instrumented region of the XFEL $\gamma\gamma$ collider.  Further, pileup has not yet been included. XCC is expected to produce, on average, nine (six) additional $\gamma\gamma\to$ hadrons interactions per bunch crossing at 380~GeV (280~GeV). Although pileup tends to degrade precision, preliminary machine-learning-based pileup-mitigation techniques suggest that we can substantially reduce its impact. A forthcoming manuscript will present a detailed evaluation of pileup and advanced algorithms for its suppression.

 \acknowledgments
The authors express their gratitude to Gudrid Moortgat-Pick, Marten Berger, and Jenny List for useful feedback on the manuscript and Caterina Vernieri for insightful discussions. Additional thanks are extended to Dimitris Ntounis for providing the \textsc{Delphes} configurations used in earlier iterations of this study. This work used the resources of the SLAC Shared Science Data Facility (S3DF) at SLAC National Accelerator Laboratory. SLAC is operated by Stanford University for the U.S. Department of Energy's Office of Science. This work is sponsored by the U.S. Department of Energy, Office of Science under Contract No. DE-AC02-76SF00515.


\appendix
\section{Pileup Mitigation with a Transformer Neural Network}
\label{sec:pileup}

This appendix describes a particle-level pileup mitigation algorithm similar to the widely-used \textsc{Puppi} algorithm \cite{Bertolini:2014bba}. The method trains a Transformer-based \cite{vaswani2023attentionneed} neural network to classify individual PFOs as originating from the hard-scatter (HS) interaction or from pileup (PU) collisions, using only the four-momentum of each constituent as input. While this procedure is not explicitly used in this analysis, we include it here as a validation study to quantify the effect of pileup on the key observables and thus event selection. After the method has been applied, we find only minimal residual differences, indicating that pileup contributions are effectively suppressed and do not substantially impact the conclusions of the main analysis. These results therefore support the treatment adopted in the baseline study, where pileup effects are neglected.

\subsection{Samples and Simulation}

\begin{table}
\centering
    \begin{tabular}{ c c c  }
\hline
\textbf{Category} & \textbf{Variable} &\quad \textbf{Definition} \quad \\
\hline
\multirow{4}{*}{Kinematics}
  & $\theta$ & The polar angle of the particle flow object. \\
& $\phi$ & The azimuthal angle of the particle \\
& $p_{T}$ & The transverse momentum of the particle \\
& $E$ & The energy of the particle \\
\hline
\end{tabular}
\caption{The set of per-particle input features used to train the set transformer model.}
\label{tab:infeats}
\end{table}

\begin{figure}
    \centering
    \includegraphics[width=0.495\linewidth]{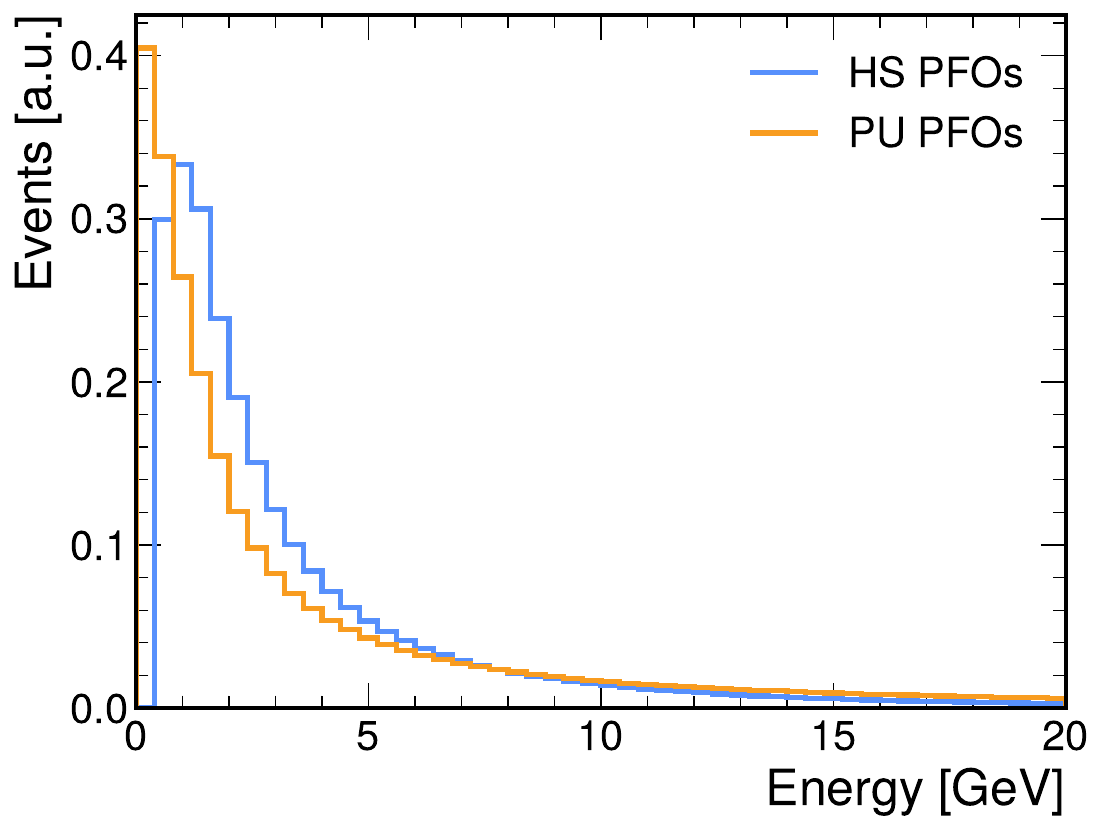}\includegraphics[width=0.495\linewidth]{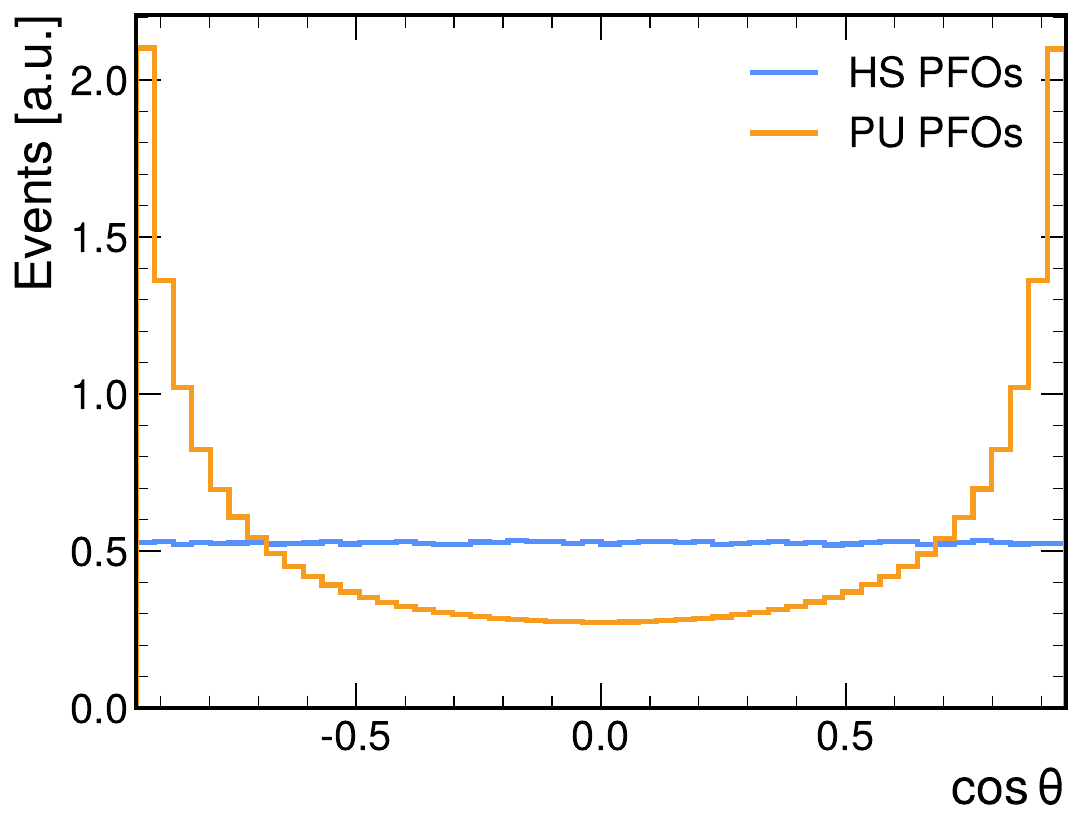}
    
    \caption{Distributions of the PFO energy (left) and $\cos \theta$ (right) for the hard-scatter (blue) and pileup (orange) PFOs.}
    \label{fig:PFO_dists_HS_PU}
\end{figure}

HS and PU events are simulated with the same \textsc{Cain}-\textsc{Whizard}-\textsc{Pythia}-\textsc{Delphes} simulation chain as described in Sec.~\ref{sec:ev_gen} of the main body. For each event, we use the full collection of PFOs, with the kinematics of each PFO used as input features, as summarized in Table~\ref{tab:infeats}. Within each event the PFOs are sorted in descending order of energy\footnote{The network is permutation-invariant and this ordering is inconsequential for performance.} and the leading $N_{\text{HS}} = 128$ PFOs are kept; events containing fewer than 128 PFOs are zero-padded. Pileup is modeled by simulating $\gamma\gamma \to \mathrm{hadrons}$ interactions in \textsc{Whizard} and is overlaid on top of the HS events. For every hard-scatter event the number of concurrent pileup interactions is sampled from a
Poisson distribution, $n_{\text{PU}} \sim \text{Poisson}(\langle\mu\rangle)$ with $\langle\mu\rangle = 9.1$ denoting the average number of interactions per bunch crossing. Further, each particle receives a binary truth label, 0 or 1, indicating HS and PU PFOs respectively. The combined event is truncated to a fixed sequence length of $N_{\text{total}} = 384$ particles, using the same zero-padding to ensure fixed-length inputs. The zeros are subsequently are ignored with masks during training and validation. We train only on $\gamma\gamma \to H$ events but evaluate the performance of our algorithm on all the processes considered in this study to ensure it robustly generalizes beyond the processes included in training. Fig.~\ref{fig:PFO_dists_HS_PU} shows the distribution of PFO energy and $\cos\theta$ distributions for the hard-scatter and pileup PFOs.

\subsection{Transformer Architecture}
The classifier network follows a Transformer encoder architecture operating on a variable-length set of PFOs within each event.  The four-dimensional kinematic input $\mathbf{x}_i$ of each particle is first mapped to a hidden representation $\mathbf{h}_i^{(0)} \in \mathbb{R}^{d}$ with $d = 128$ through a learned linear projection. The embedded sequence is then processed by a stack of $L = 6$ identical Transformer encoder layers, each comprising multi-head self-attention with $n_H = 8$ heads followed by a feed-forward network of inner dimension $d = 512$, with dropout at rate $p = 0.1$ and ReLU activation applied within each layer. The self-attention mechanism allows every particle to exchange information with all other constituents in the event, enabling the network to exploit the collective kinematic structure of both the hard scatter and pileup.  The contextualised representation $\mathbf{h}_i^{(L)}$ produced by the final encoder layer is passed through a per-particle classifier head consisting of a two-layer MLP ($d, d/2, 1$) with ReLU activation and dropout (with $p = 0.1$). The final output is a scalar logit whose sigmoid gives the predicted pileup probability for each PFO $\widehat{y}_i \in [0, 1]$.

The mixed dataset is split into training (70\%), validation (10\%), and test (10\%) subsets, and zero-padded particle slots are masked and excluded from all loss and metric computations. The network is trained for 30 epochs to minimize the focal cross-entropy loss function:
\begin{equation}
  \mathcal{L}_{\text{focal}}
    = -\frac{1}{N_{\text{valid}}}
      \sum_{i=1}^{N_{\text{valid}}}
      (1 - p_{t,i})^{\gamma}\,
      \bigl[-y_i \log \widehat{y}_i - (1 - y_i)\log(1 - \widehat{y}_i)\bigr]\,,
  \label{eq:focal_loss}
\end{equation}
where $p_{t,i} = \widehat{y}_i$ if $y_i = 1$ and $p_{t,i} = 1 - \widehat{y}_i$ otherwise, and $\gamma = 2.0$ is the focusing parameter.  The focal loss down-weights the contribution of well-classified particles and concentrates learning on the more ambiguous boundary between hard-scatter and pileup constituents. This also has the advantage of dynamically mitigating the class imbalance that arises when the number of pileup particles differs substantially from the number of hard-scatter particles in a given event.  The \textsc{AdamW} \cite{j.2018on} optimizer is chosen with at an initial learning rate of $5 \times 10^{-4}$, decayed to $10^{-6}$ via a cosine-annealing \cite{loshchilov2017sgdrstochasticgradientdescent} schedule, with a batch size of 128 events. The model checkpoint with the lowest validation set loss is used for all subsequent inference on the test set.

\subsection{Results}

\begin{table*}[tb]
\centering
\begin{tabularx}{\linewidth}{@{} l *{6}{>{\centering\arraybackslash}X} @{}}
\toprule
& \multicolumn{3}{c}{\textbf{Energy Resolution}} & \multicolumn{3}{c}{\textbf{Mass Resolution}}\\ \midrule
& Mean & Median & IQR & Mean &  Median &  IQR \\
\midrule
$\boldsymbol{\gamma\gamma \to HH \to b\overline{b}b\overline{b}}$
  & 0.017 & 0.000 & 0.027 & 0.644 & 0.000 & 0.107 \\ \midrule
$\gamma\gamma \to W^+W^- \to q\overline{q}q\overline{q}$
  & 0.027 & 0.000 & 0.042 & 0.719 & 0.000 & 0.183 \\
$\gamma\gamma \to t\overline{t}$
  & 0.008 & 0.000 & 0.028 & 0.567 & 0.000 & 0.105 \\
$\gamma\gamma \to ZZ \to q\overline{q}q\overline{q}$
  & 0.021 & 0.000 & 0.021 & 0.156 & 0.000 & 0.090 \\
$\gamma\gamma \to q\overline{q}$
  & 0.024 & 0.000 & 0.051 & 0.213 & 0.000 & 0.239 \\
$\gamma\gamma \to ZH \to q\overline{q}H$
  & 0.017 & 0.000 & 0.020 & 0.535 & 0.000 & 0.086 \\
\midrule
$e\gamma \to eq\overline{q}$ or $\nu q\overline{q}$
  & 0.025 & 0.000 & 0.086 & 0.715 & 0.000 & 0.366 \\
$e\gamma \to eq\overline{q}q\overline{q}$ or $\nu q\overline{q}q\overline{q}$
  & 0.021 & 0.000 & 0.042 & 0.463 & 0.000 & 0.175 \\
$e\gamma \to eq\overline{q}H$
  & 0.021 & 0.000 & 0.037 & 0.229 & 0.000 & 0.159 \\
\midrule
$e^+e^- \to q\overline{q}$
  & 0.024 & 0.000 & 0.031 & 0.223 & 0.000 & 0.174 \\
$e^+e^- \to W^+W^-/ZZ \to q\overline{q}q\overline{q}$
  & 0.031 & 0.000 & 0.043 & 0.298 & 0.000 & 0.187 \\
$e^+e^- \to ZH \to q\overline{q}H$
  & 0.021 & 0.000 & 0.032 & 0.979 & 0.000 & 0.132 \\
$e^+e^- \to t\overline{t}$
  & 0.010 & 0.000 & 0.024 & 0.608 & 0.000 & 0.081 \\
\bottomrule
\end{tabularx}
\vspace{7.5pt}
\caption{Per-process jet energy and mass resolution of the Transformer-based
pileup classifier, evaluated on the held-out test set.  The resolutions are
defined as $\Delta E/E = (E_{\mathrm{pred}} - E_{\mathrm{truth}}) /
E_{\mathrm{truth}}$ and $\Delta M/M = (M_{\mathrm{pred}} -
M_{\mathrm{truth}}) / M_{\mathrm{truth}}$, where the prediction uses the
working point $\widehat{y}_i < 0.5$. 
\label{tab:perfpertop}}
\end{table*}


\begin{figure}
    \centering
    \begin{minipage}{0.495\linewidth}
    \centering
        \includegraphics[width=\linewidth]{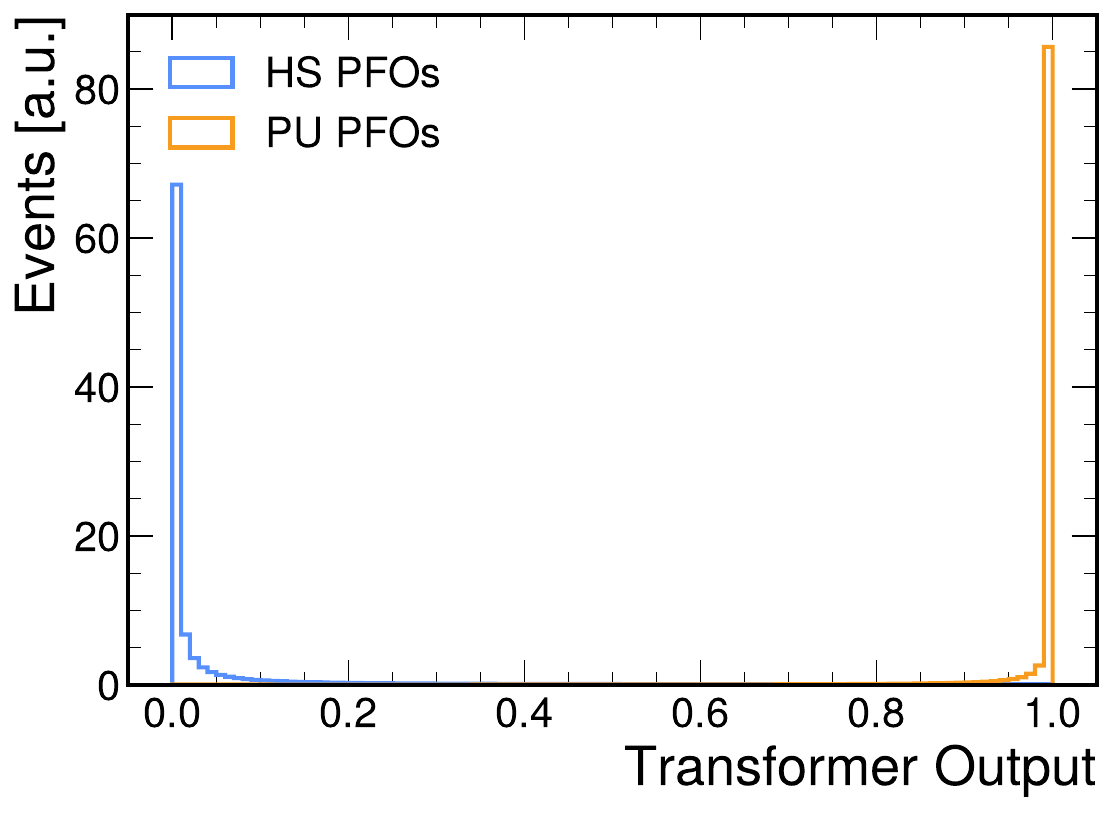}
    \end{minipage}
    \begin{minipage}{0.495\linewidth}
        \includegraphics[width=\linewidth]{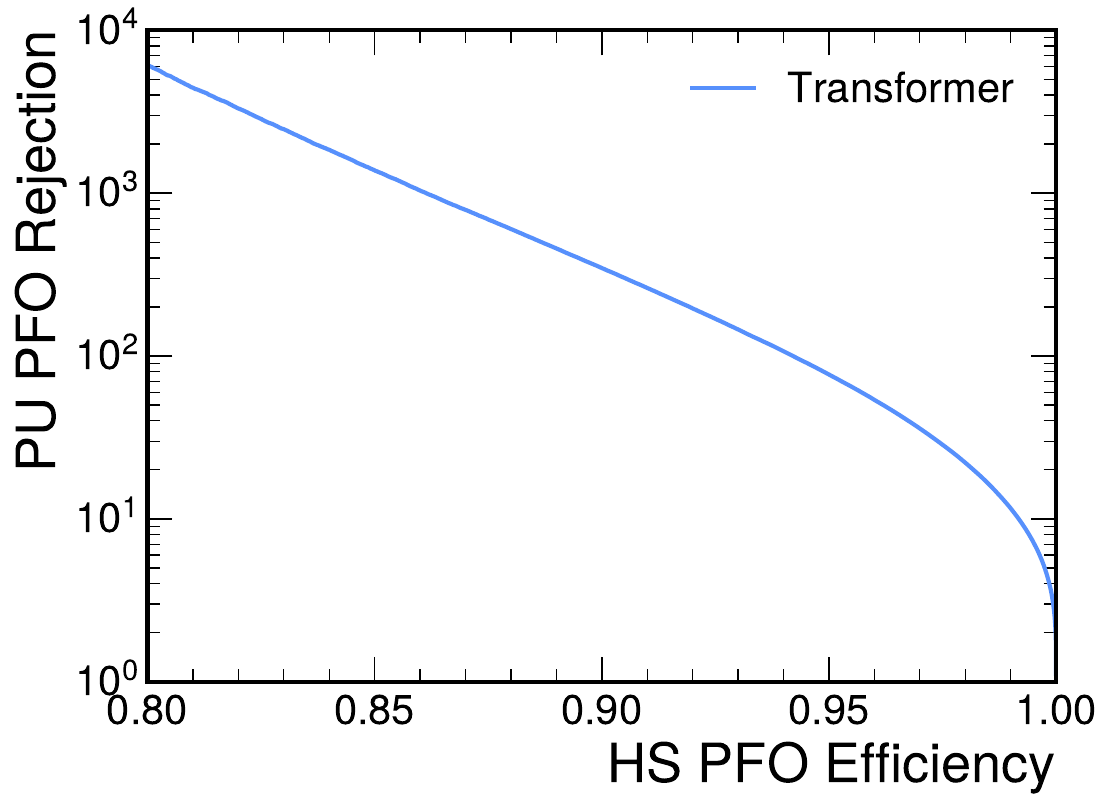}
    \end{minipage}
    \caption{The transformer output distributions for the HS PFOs and PU PFOs (left), and the ROC curve showing the PU PFO rejection as a function of the HS PFO efficiency (right).}
    \label{fig:perf_pu_transformer}
\end{figure}
The performance of our method is evaluated on the held-out test set. Fig.~\ref{fig:perf_pu_transformer} shows the neural network output distributions for the HS and PU PFOs as well the PU PFO rejection as a function of the HS PFO efficiency. From the plots, we observe that the method achieves excellent separation of the PU and HS particles.   

\begin{figure}
    \centering
    \begin{minipage}{0.495\linewidth}
        \includegraphics[width=\linewidth]{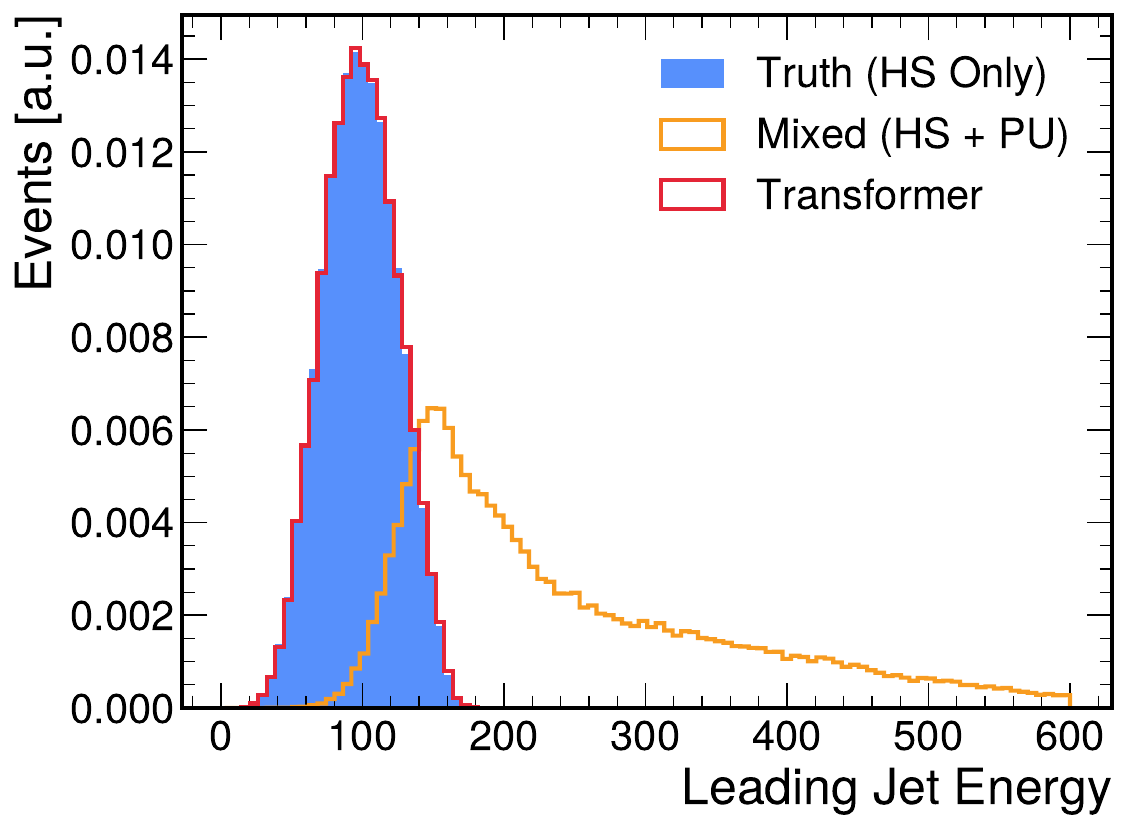}
    \end{minipage}
    \begin{minipage}{0.495\linewidth}
        \includegraphics[width=\linewidth]{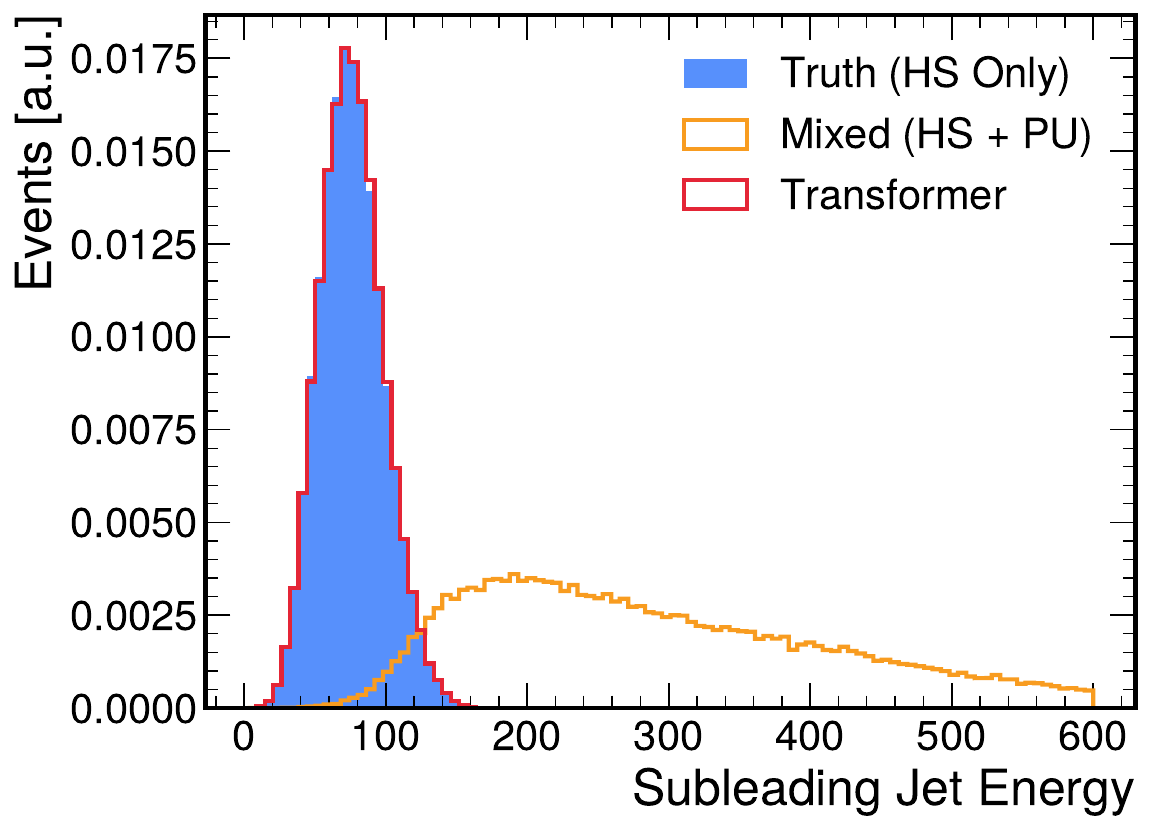}
    \end{minipage}
    \begin{minipage}{0.495\linewidth}
        \includegraphics[width=\linewidth]{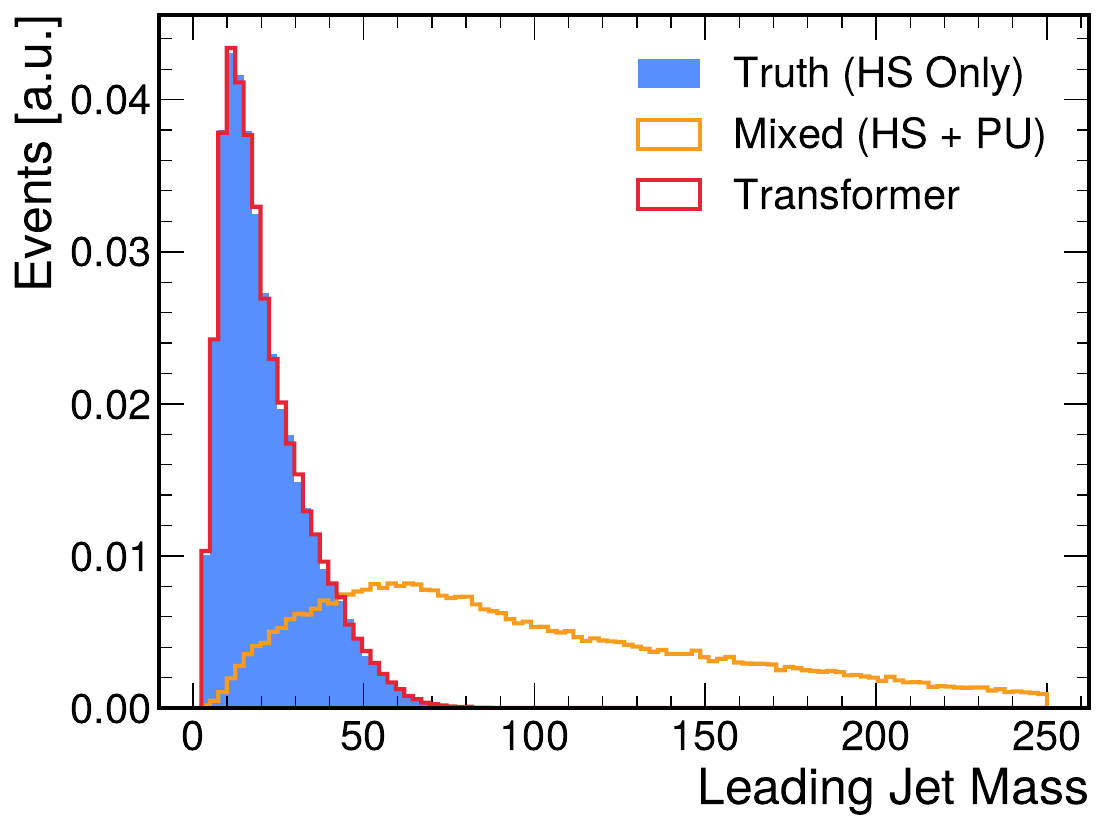}
    \end{minipage}
    \begin{minipage}{0.495\linewidth}
        \includegraphics[width=\linewidth]{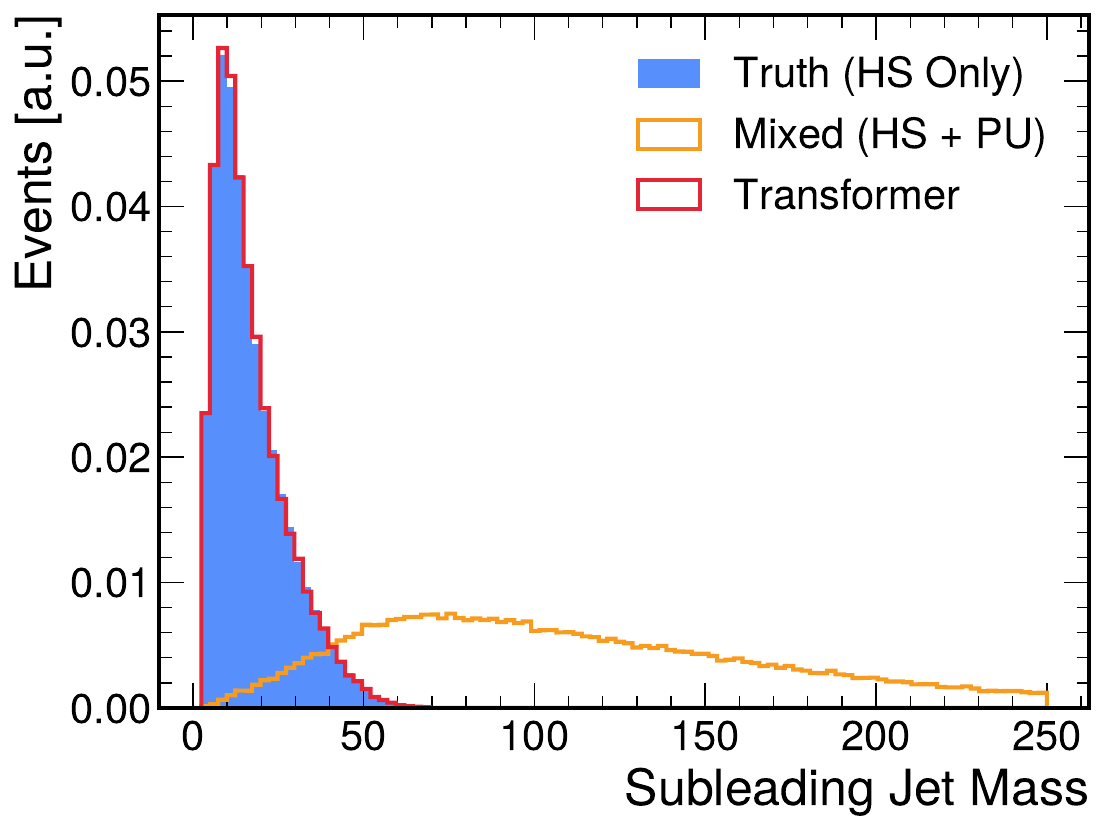}
    \end{minipage}
    \caption{Distributions of the leading jet energy (top-left), subleading jet energy (top-right), leading jet mass (bottom-left), and subleading jet mass (bottom-right) for the ground truth (blue filled), mixed events (orange), and the transformer-based pileup subtracted events (red).}
    \label{fig:phys_obs}
\end{figure}

\begin{figure}
    \centering
    \begin{minipage}{0.495\linewidth}
        \includegraphics[width=\linewidth]{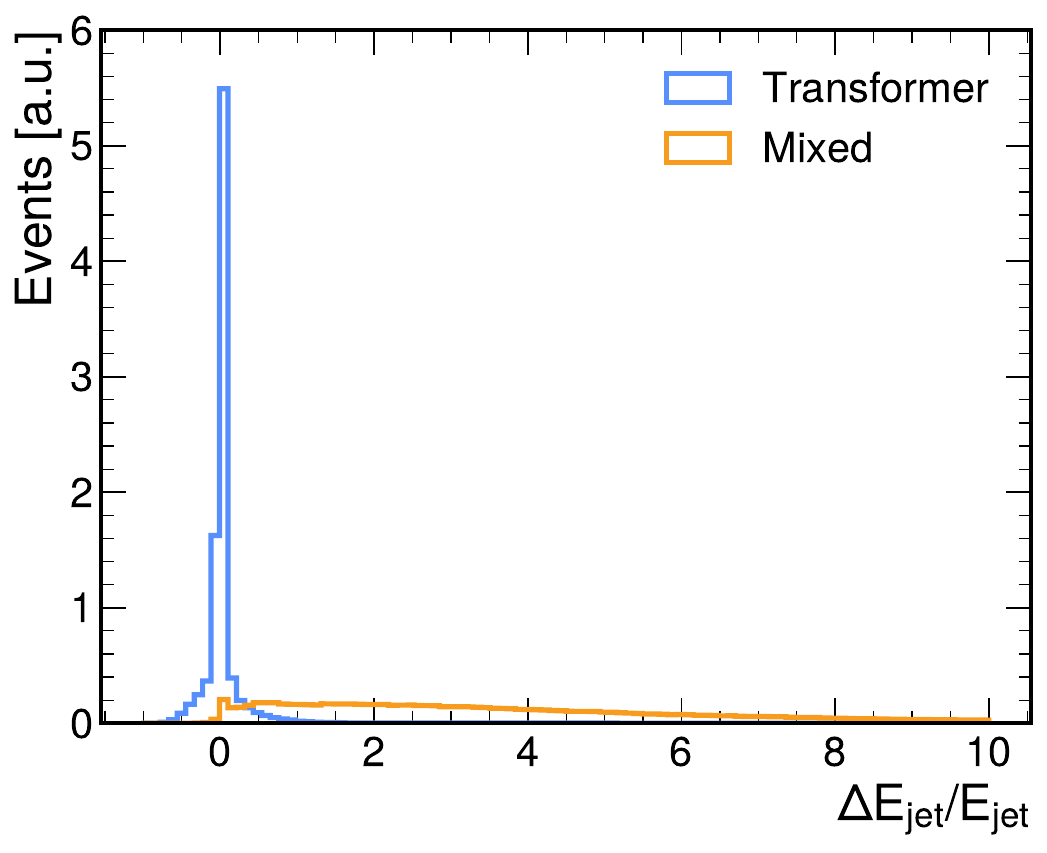}
    \end{minipage}
    \begin{minipage}{0.495\linewidth}
        \includegraphics[width=\linewidth]{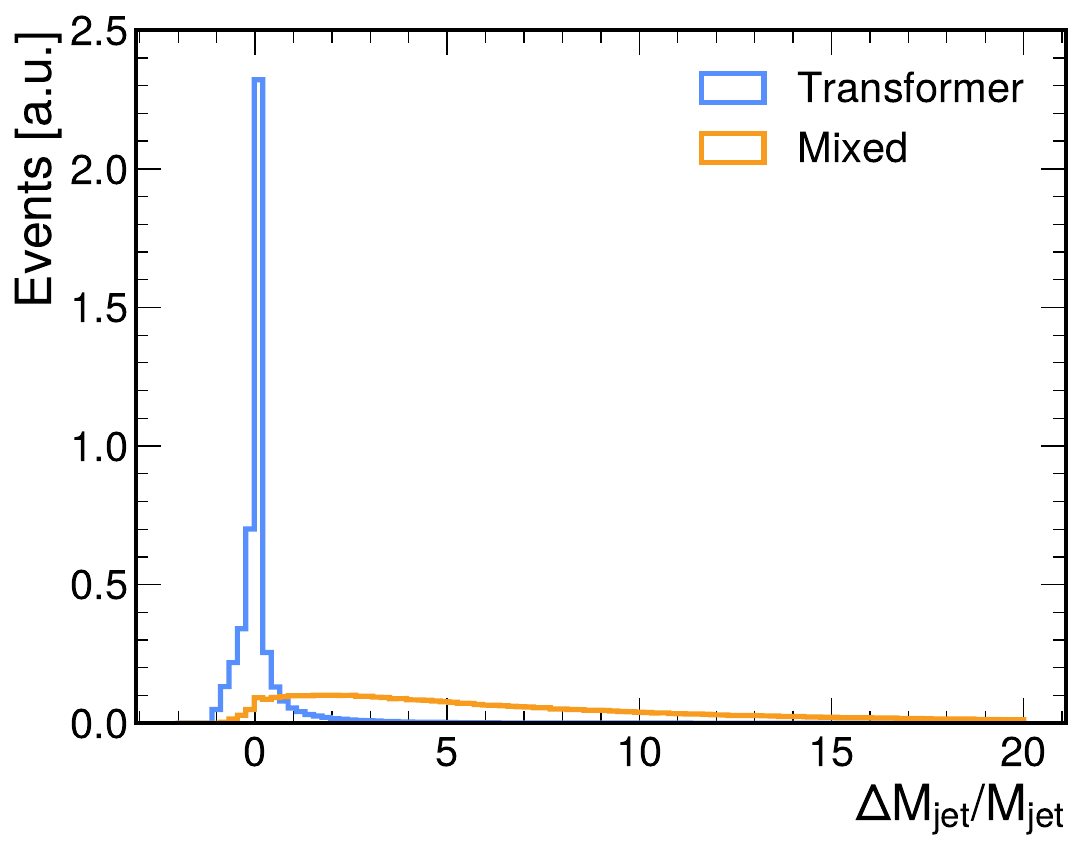}
    \end{minipage}

    \caption{Distributions of the jet energy resolution (left), and jet mass resolution (right) for the events after transformer-based subtraction (blue), and mixed events (orange).}
    \label{fig:phys_obs_jes_jms}
\end{figure}

To evaluate the physics performance of our pileup mitigation technique, we compute several jet observables and compare the ground truth values of the observables to the reconstructed values obtained after pileup subtraction is applied. We use a working point defined by a threshold of $\widehat{y}_i > 0.5$ for the binary classification decision i.e.~PFOs with $\widehat{y}_i > 0.5$ are classified as PU and removed. The surviving particles are then clustered using the Durham algorithm with $n_\mathrm{jets}=4$ as before. We define the (sub-)leading jet as the one with the (second-)highest energy, and show the distributions of their energies and masses for the ground truth, the mixed events, and our transformer-based method in Fig.~\ref{fig:phys_obs}.  While these plots are standard and provide a leading-order estimate of performance, a more useful metric is the jet energy (mass) resolution defined as $\Delta E / E = (E_\mathrm{reco}-E_\mathrm{truth})/E_\mathrm{truth}$ ($\Delta M / M = (M_\mathrm{reco}-M_\mathrm{truth})/M_\mathrm{truth}$) shown in the left (right) hand panel of Fig.~\ref{fig:phys_obs_jes_jms} for all four jets in the event. The transformer prediction is sharply peaked around 0, with $X\%$ of all events contained within a deviation of $<1\%$. Table \ref{tab:perfpertop} shows the means, medians, and inter-quartile ranges (IQR) of the jet mass and energy resolutions for all the signal and background events considered in this study. All processes have means $<0.03$ for the jet energy resolution and vanishing medians for both the mass and energy resolutions.  

\bibliographystyle{JHEP}
\bibliography{biblio.bib}


\end{document}